\documentclass{article}

\usepackage{pgf,tikz}
\usetikzlibrary{arrows}
\usepackage{array,multirow,makecell}
\setcellgapes{1pt} \makegapedcells
\usepackage{amssymb}
\usepackage{subfig}
\usepackage[normalem]{ ulem }
\usepackage{soul}
\usepackage{graphicx}
\usepackage{here}
\usepackage{epsfig}
\usepackage{color}
\usepackage{pgf,tikz,amsthm}
\usetikzlibrary{arrows}
\usepackage{amsmath}
\usepackage{epstopdf}

\usepackage{breqn}

\usepackage{hyperref}
\hypersetup{
	colorlinks=true,
	linkcolor=blue,
	filecolor=magenta,      
	urlcolor=cyan,
	citecolor=red
}
\usepackage[capitalise]{cleveref}

\newtheorem{theorem}{\textbf{Theorem}}

\newtheorem{definition}{\textbf{Definition}}

\newcommand{\comment}[1]{}

\usepackage{tikz}
\usepackage{anyfontsize}

\title{A multi-patch model for sylvatic hematophagous vector dynamics: the impact of landscape anthropization}
\author{I.V. Yatat-Djeumen$^{1,2,3}$\footnote{Corresponding author: valaire.yatat@univ-yaounde1.cm}, L. Eigentler$^{4,5}$, O.W. Happi-Tchakount\'e$^{1}$, I. Tcheferi$^{6,7}$,  P. Couteron$^{2}$ \\	
	\small $^1$University of Yaound\'e I, National Advanced School of Engineering of Yaound\'e, \\ 
	\small Department of Mathematics and Physics, Yaound\'e, Cameroon \\
    \small $^2$AMAP, Univ Montpellier, IRD, CIRAD, CNRS, INRA Montpellier, France \\
	\small $^3$IRD, Sorbonne Université, UMMISCO, F-93143, Bobby, France \\ 
    \small $^4$ Warwick Mathematics Institute, University of Warwick, Coventry CV4 7AL, United Kingdom \\
    \small $^5$ Zeeman Institute for Systems Biology \& Infectious Disease Epidemiology Research, \\ \small University of Warwick, Coventry CV4 7AL, United Kingdom \\
	\small $^6$National Institute of Cartography, Yaoundé, Cameroon \\
	\small $^7$University of Yaoundé I, Teacher Training College, Department of Biology, \\
	\small Plant Systematics and Ecology Laboratory,	   Yaoundé, Cameroon\\
}
\date{\today}

\begin{document}

\maketitle

\begin{abstract}
	In this paper, we develop and study a multi-patch model for sylvatic hematophagous vector--(wild) animal interactions with nonlinear migration terms that depend on landscape anthropization.
    It is assumed that any patch is a suitable habitat for both animals and vectors. However, anthropization of a patch was assumed to both reduce the animal carrying capacity and to induce animal and vector displacement. We considered three types of dispersal. Under unconstrained global dispersal, displaced individuals can move to any other patch. Under the more realistic random and constrained neighbor dispersal, movement is restricted to neighboring patches, and displaced animals choose a neighboring patch either at random or preferentially among the less anthropized ones.  
    We parameterized our model for a population of tsetse flies (\textit{Glossina} spp.) in  the districts of Campo, Ny\'e'\'et\'e, and Ma'an, Ocean division, Cameroon. We characterized the equilibria and their local stability analytically for unconstrained global dispersal, and found numerically that, under both neighbor dispersal regimes, animal and vector populations progressively expand before ultimately settling into a stable equilibrium. We found that, under random neighbor dispersal, the impact of anthropization on the migration dynamics had a much larger effect on the steady state densities than the impact of anthropization on animal carrying capacity. In the constrained neighbor dispersal setting, the rate of population spreading was noticeably slower than in the case of random neighbor dispersal, and numerous patches maintained extremely sparse population levels. We further investigated how different anthropization scenarios affected the speed of invasion. We found that for random neighbor dispersal, invasion times decreased upon introduction of newly anthropized regions, while in the case of constrained neighbor dispersal, anthropized regions either increased invasion times or had a neutral effect. Approximating invasion dynamics by pulled traveling waves, we found that anthropized regions act as ``invasion highways'' or ``invasion barriers'' depending on the dispersal type. Ecologically, we conclude that more empirical data on animal dispersal are needed to better understand these population dynamics.
\end{abstract}

\section{Introduction}

Across Africa, the landscape is undergoing persistent transformation, primarily characterized by the expansion of cultivated land and human settlements, coupled with the extensive loss and degradation of natural woody vegetation. This shift is resulting in a growing human utilization of the ecosystem's productivity. Recent decades have seen accelerated change in regional hotspots, where a measurable decline in natural cover corresponds to significant increases in built-up areas and cropland \cite{Bullock2021-vt,Chiaka2021-hw,Fetzel2016-xc,Kimani2021-jb}. For instance, by 2005, the continent-wide human consumption of net primary production had reached approximately 20\%, marking a roughly 55\% increase since 1980. This intensity and efficiency of biomass use varies considerably across regions \cite{Kimani2021-jb}. According to \cite{Bullock2021-vt}, cropland area in a seven-country East Africa sample increased by approximately 18,154,000~ha ($\approx 34.8\%$) between 1998 and 2017, and settlement area rose by $\approx 43.5\%$ over the same period; woody classes were largely converted to less-woody uses with approximately 20 million hectares more conversion than succession toward woody recovery. In the same vein, \cite{land11060850} reported that natural cover around major mining agglomerations in southeastern Katanga (Democratic Republic of the Congo) lost $>$60\% of its area between 1979 and 2020, with agricultural and built-up areas increasing sharply. Moreover, it was estimated that in Ivory Coast, dense forest and degraded forest in the south‑west declined at rates of about 1.44\%/yr and 3.44\%/yr respectively from 1987 to 2015, driven primarily by conversion to agriculture \cite{Chiaka2021-hw}.

Landscape (or spatial) epidemiology is an interdisciplinary field that examines the spatial distribution and dynamics of infectious diseases in relation to environmental factors and landscape structure \cite{Emmanuel2011,Lambin2010,Ostfeld2005,Reisen2010}. This approach integrates ecological, geographical, and epidemiological knowledge to identify how landscape features, such as land use, habitat fragmentation and climate, influence the spread of diseases across different regions. According to Pavlovsky, the dynamics of a transmissible disease rely on permanent interaction of the quintet of its prerequisites, that are (1) animal donors; (2) vectors; (3) animal recipients; (4) the pathogenic agent itself in an infective state, (5) the influence of factors of the external environments contributing to the circulation of pathogenic agent (\textit{in} \cite{Emmanuel2011,Lambin2010}). Spatial interactions between these agents in a landscape explain patterns of infection risk and may contribute to disease emergence.

The emergence and re-emergence of vector-borne and zoonotic diseases are influenced by ecosystem changes at the landscape level, along with other factors \cite{Lambin2010,Orta-Pineda2021-yr}. According to Patz et al. \cite{Patz2004}, anthropogenic land use modifications, encompassing a spectrum of activities from deforestation,  road construction, agricultural encroachment to urban expansion, represent primary drivers of emerging infectious disease events and exert substantial influence on the transmission dynamics of endemic infections, see also \cite{Emmanuel2011,Orta-Pineda2021-yr,Reisen2010}. These modifications initiate a cascade of secondary effects, including habitat fragmentation, pathogen introduction, environmental pollution, socioeconomic disparities and human migration, which collectively exacerbate the risk of infectious disease emergence \cite{Patz2004}. More precisely, the dynamics of landscapes may significantly influence the movement patterns of vectors and animal hosts, which in turn affect disease transmission patterns \cite{Crowder2013,Crump2021-ze,Ferraguti2021,Fletcher2018,Lambin2010,Orta-Pineda2021-yr,Shaw2024,Tracey2014,Vale2024-lb}. Various studies highlight how landscape attributes, such as fragmentation and connectivity, shape the interactions between hosts, vectors, and pathogens. Indeed, habitat fragmentation can limit animal movement, affecting the spread of diseases. 

Similarly, Crowder et al. \cite{Crowder2013} and Ferraguti et al. \cite{Ferraguti2021} highlighted the impact of landscape dynamics on mosquito communities and their dynamics, which in turn have an impact on the transmission dynamics of pathogens like West Nile virus and avian malaria. 
 Indeed, landscape structure may alter the distribution, abundance, and community composition of mosquito vectors and avian hosts, concentrating pathogen transmission risks in modified or wetland interfaces, see also \cite{Froxan-Grabalosa2025-lh}. Furthermore, anthropogenic environmental disruptions, particularly land-use changes and agricultural expansion, may play a critical role in the appearance of mosquito-borne pathogens within wild ecosystems (see e.g., \cite{Ferraguti2016-lq,Orta-Pineda2021-yr,Perrin2022-iz,Perrin2023-xg} and references therein). Specifically, the alteration of natural habitats and the resulting decline in wildlife populations significantly restrict the availability of hosts for numerous mosquito species. An agent-based model demonstrated that intermediate levels of habitat permeability can enhance disease transmission among wildlife populations \cite{Tracey2014}. Connectivity between habitats may also be a key component of the disease transmission pattern. Conversely, fragmented landscapes can lead to increased tick density and pathogen prevalence, as shown in studies of tick-borne diseases in the United States \cite{Shaw2024}.

Numerous contributions used mathematical modeling to elucidate the spatio-temporal dynamics of hosts and vectors. Common approaches include reaction-diffusion-advection models, patch or metapopulation models, agent-based or stochastic simulations, and hybrid approaches (e.g., \cite{Cailly2011,Cui2025-bt,daSilva2020,Dufourd2012-dd,Dufourd2013-uu,dumont_spatio-temporal_2011, Dye2024-sw, Hancock2019, Lutambi2013, Lutambi2013MBS, Maneerat2016-th,McCormack2019,nguyen:tel-04687836, Roques2016,Tracey2014} and references therein). Dumont and Dufourd \cite{dumont_spatio-temporal_2011} considered a dispersal model featuring only adult female mosquitoes split into blood meal searchers and breeding site searchers. This work was then extended by Dufourd and Dumont \cite{Dufourd2012-dd}, who took into account the aquatic stage, immature females, resting females, wild males and sterile males and designed a reaction-advection-diffusion model for mosquito dynamics. They further considered heterogeneity in the landscape by assuming that areas more favorable to mosquitoes were characterized by lower diffusion. By relying on numerical simulations of the resulting reaction-advection-diffusion model, they showed that landscapes can have a strong influence on mosquito distribution and on the efficiency of vector control tools.  Da Silva et al. \cite{daSilva2020} model the spatial population dynamics of the \textit{Aedes aegypti} mosquito using a reaction-diffusion partial differential equation model that takes account of insecticide application. The model considers mosquitoes in the mobile, aquatic and egg phases. Their study suggests that a weekly insecticide application is optimal for reducing the mosquito population and that using models based on partial differential equations is an effective tool for simulating and analyzing control strategies. Dye and Cain \cite{Dye2024-sw} used a mathematical model that simulates mosquito populations in distinct habitats and calculates the migration rates between them with the aim of predicting the conditions under which the release of \textit{Wolbachia} carriers can lead to their sustained presence near the release site or even spread to other territories. The model predicts three possible outcomes for a single release, namely failure to establish, establishment in the release area only or establishment in multiple areas. The paper of Roques and Bonnefon \cite{Roques2016} proposed a mechanistic-statistical framework to model population dynamics in landscapes with linear elements like roads and rivers. It used a hybrid system of coupled two-dimensional (landscapes) and one-dimensional (corridors, i.e., roads, hedgerows, rivers) reaction-diffusion equations to describe population changes in the 2D matrix and the 1D corridors. The results showed that the corridors play an important role in species dynamics. The work of Hancock et al. \cite{Hancock2019} dealt with an age-structured model of the spatial dynamics of \textit{Wolbachia} in \textit{Aedes aegypti} populations, which incorporates features of mosquito demographic variation related to landscape heterogeneity that affect \textit{Wolbachia} spread, such as density dependence. The multi-patch model takes into account juvenile stages, male and female adults of mosquitoes. They assumed that density-dependent relationships differ across subpopulations, and constructed heterogeneous landscapes for which model-predicted patterns of variation in mosquito abundance and demography approximate those observed in field populations. They showed that their model can produce rates of spatial spread of \textit{Wolbachia} similar to those observed in field releases and can be applied to compare the performance of different strategies. Lutambi et al. \cite{Lutambi2013,Lutambi2013MBS} proposed a model coupling spatial and temporal dynamics of mosquitoes, in order to identify the environmental parameters influencing the abundance and dispersal of mosquitoes and to identify those that are most influential. Their  results showed that the landscape heterogeneity and its evolution over time influence the distribution and abundance of mosquitoes. They relied on a discrete-space continuous-time or multi-patch mathematical model to examine the impact of mosquito dispersal and the heterogeneous distribution of resources (such as hosts and breeding sites) on the spatial distribution and dynamics of mosquito populations. The findings highlight the importance of considering environmental heterogeneity and resource availability when developing effective control strategies. A spatially explicit
agent-based simulation model of \textit{Aedes aegypti} mosquito was developed to explore its population dynamics in urban areas  \cite{Maneerat2016-th}. This model focuses on the female \textit{Aedes aegypti} mosquito and simulates adult mosquitoes as agents that interact with their local environment, which provides resources and constraints for their development and behavior. Simulation results showed a significant relationship between urban topology, human densities and adult mosquito flight, dispersion and egg-laying behaviors.

However, to the best of our knowledge, landscape metrics and (wild) animal dynamics were often ignored or only implicitly considered in the available mathematical models designed for the study of (hematophagous) vector dynamics. In addition, most of them relied almost exclusively on numerical simulations. We acknowledge the existence of several contributions dealing with the spatio-temporal dynamics of vector-borne epidemiological models where in addition to the vectors, human and/or animal populations are also considered together with an infectious disease, see for example \cite{Anzo-Hernandez2018-to,Arino2012-mp,Auger2008-lc,Esteva2019-mf,Gimenez-Mujica2025-sf,Iggidr2017-jj,Moulay2013-uu} and references therein. However, even in these works, landscape dynamics are scarcely considered. We do not consider these models further, since both humans and infectious diseases are beyond the scope of the present contribution.

The present contribution aims to explicitly combine landscape structure with hematophagous, sylvatic vectors (e.g. mosquitoes, tsetse flies) and (wild) animal dynamics by using mathematical modeling and advocating mathematically tractable models. The present contribution is the first step of a broader project investigating the impact of landscape dynamics on the emergence of zoonotic infectious diseases. It aims to understand interactions between  hematophagous, sylvatic vectors and wild animals in a heterogeneous landscape. The present contribution extends the two-patch model that we developed and studied in \cite{Happi2026} to a multi-patch model. Moreover, we also aim to characterize the impact of some anthropization scenarios on the dynamics of both hematophagous, sylvatic vectors and (wild) animals. The next step could consist in considering, in addition to the two previous populations, both human dynamics together with a vector-borne infectious disease that may allow spillover from wild animals to the human population.

The remainder of the paper is organized as follows: Section \ref{section:study-area} presents the study area together with the characterization of the related levels of anthropization. Section \ref{section:mathematical-models} deals with the multi-patch mathematical model that explicitly takes into account anthropization-driven migration between patches. Section \ref{section:numerical-simulations} focuses on numerical simulations of the model taking into account several anthropization scenarios. Section \ref{section:conclusion} summarizes and discusses our results. Background results on matrices and additional figures are given in the appendices.

\section{Study area and anthropization characterization}\label{section:study-area}

\subsection{Description of the study area}
\begin{figure}
    \centering
    \includegraphics[width=0.99\linewidth]{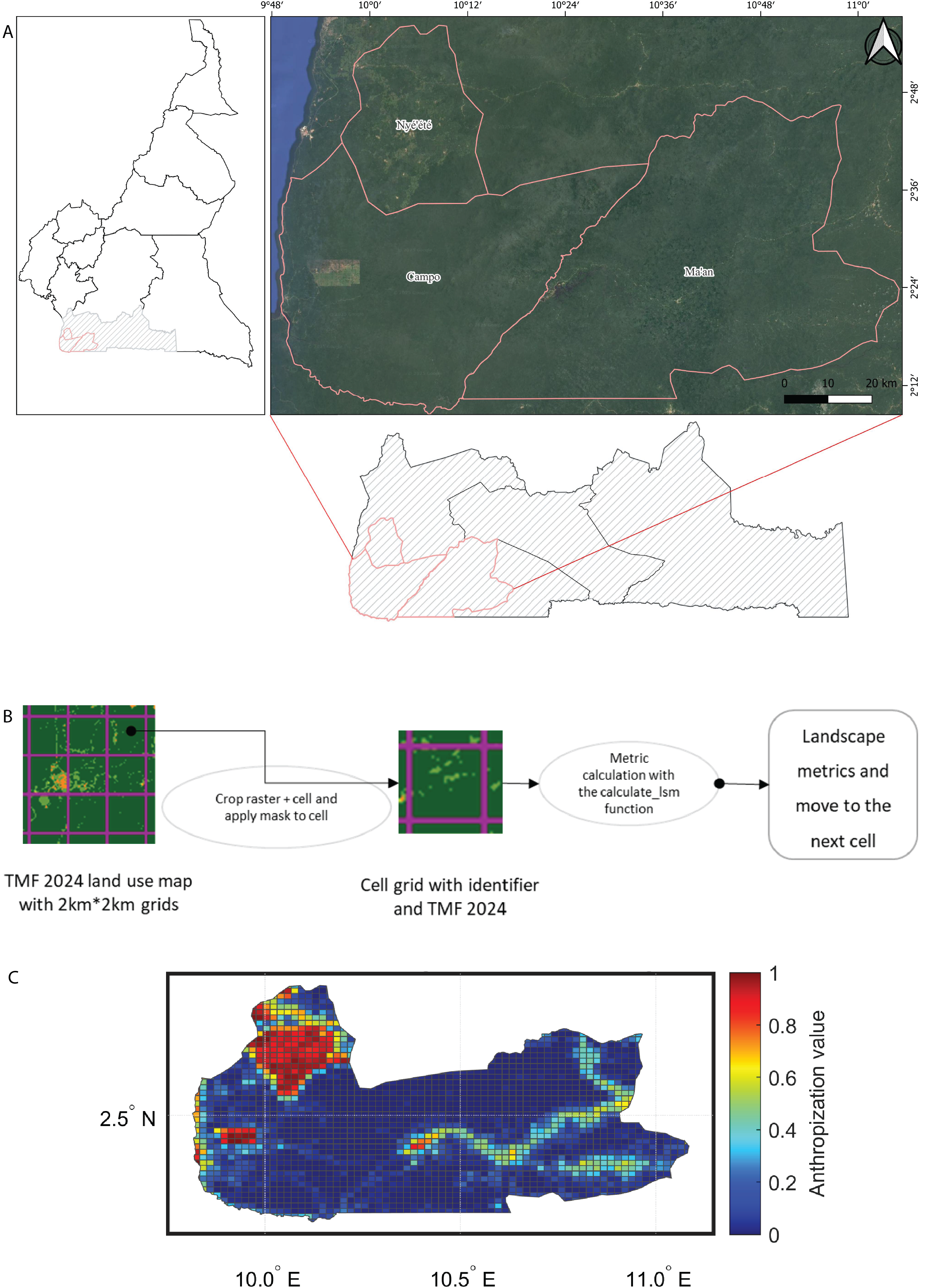}
    \caption{\textbf{A:} Study area covering the districts of Ma'an, Campo, and Ny\'e'\'et\'e in southern Cameroon. \textbf{B:} Methodological workflow used for the landscape metrics extraction. \textbf{C:} Spatial distribution of the anthropization levels on a 2~km $\times$ 2~km spatial grid.}
    \label{fig: fig1}
\end{figure}

The study was conducted in the districts of Campo, Ny\'e'\'et\'e, and Ma'an (\cref{fig: fig1}A), located in the Ocean division (South Region, Cameroon), naturally covered by dense rainforest (Guineo-Congolian region of the Biafran forest type, where the Caesalpiniaceae family dominates  \cite{letouzey1985notice}). The climate is equatorial, very hot and humid, with annual rainfall between 2500 and 4000 mm and a short dry season. Temperatures range between 22 and 29$^\circ$C. These climatic conditions, combined with ferralitic soils, are deemed favorable for the development of oil palms  \cite{Olivry1986-es}. Yet, palm tree plantations are still modest with the notable exception of the CAMVERT plantation.  This area is also home to the Campo Ma'an National Park, located on the border between Cameroon and Equatorial Guinea.

\subsection{Annual map of forest cover change}
Global data based on Landsat image archives provide historical information on land cover with a resolution of 30~m, particularly with regard to forest cover, thus enabling the assessment of disturbances and deforestation. Based on this historical series, annual maps of land cover changes have been derived and are accessible in the tropical moist forest (TMF) databases (see for instance \cite{Vancutsem2021-ac}), which provide annual maps of land cover changes from 1990 to 2024, distinguishing between degradation and deforestation of forest cover. According to \cite{GFWSudCameroon2024}, from 2002 to 2024, the Ocean division lost 9\% of its forest area, of which 65\%, i.e. $ca.$ 85,000~ha, were initially moist undisturbed forest.

\subsection{Creation of spatial grids and extraction of landscape metrics}

For this work, we downloaded the annual map of forest cover changes for the year 2024 from the TMF portal (https://forobs.jrc.ec.europa.eu/TMF/data). These data layers feature land use/cover classes with a spatial resolution of 30 m inherited from the Landsat historical series. In order to get a spatial representation of the proportion of each land use/cover type, we generated a 2~km by 2~km spatial grid covering the area of interest (see \cref{fig: fig1}C). An R function was written to create and delineate the raster of patch identifiers, then calculate landscape metrics using the \textit{calculate$\_$lsm} function from the landscapemetrics package \cite{Hesselbarthetal2019}. The function was then applied to all patches to obtain landscape variables across the entire grid. Among the available metrics, we selected \textit{cpland}, which corresponds to the proportions of forest cover types (i.e., `Undisturbed' tropical moist forest (TMF), degraded TMF and TMF regrowth) in the area of interest (see \cref{fig: fig1}B). The proportions were normalized between 0 and 1, see \cref{fig: fig1}C.



\subsection{The anthropization indices}

Noting that the default land cover of each patch, in the absence of severe disturbance, is forest, 
we compute the normalized indices $\alpha_i\in[0,1)$ of each patch $i\in\{1,2,...,n\}$, which correspond to the level of vegetation cover loss resulting from anthropogenic pressure related to urbanization and deforestation.
We assume that for a given patch $i\in\{1,2,...,n\}$:
\begin{equation}\label{alpha:formule-de-calcul}
	\alpha_i=1-(\text{Undisturbed tropical moist forest + Degraded tropical moist forest + Tropical moist forest regrowth}).
\end{equation}


The computed values for every patch are summarized in \cref{fig: fig1}C.


\section{Mathematical model}\label{section:mathematical-models}
In this section, we develop a multi-patch mathematical model for sylvatic hematophagous vector dynamics. The model accounts for the vital dynamics (i.e., birth and death) of sylvatic hematophagous vectors and of their hosts (wild animals, which provide blood meals), as well as their dispersal. 

We divide the two-dimensional space into a set of discrete patches (see \cref{fig: fig1}C), each corresponding to a $2~\mathrm{km} \times 2~\mathrm{km}$ area \cite{Vale1984}. 

\subsection{The $n$-patch model with unconstrained global dispersal}\label{sec:global}	
In this section, we consider the population model composed of $n$ patches interconnected through nonlinear dispersal dynamics and where every patch $i$ is occupied by both hematophagous vectors ($V_i$) and wild animals ($A_i$). Our objective is to understand the impact of anthropization, viewed through the lens of natural landscape fragmentation at large scale and inter-patch connectivity at local scale, on the dynamics of both sylvatic hematophagous vectors and wild animal population. To model this, we introduce $\alpha_{i}\in[0,1)$ as a measure of anthropization pressure on patch $i$. An $\alpha_{i}$ 
value of 0 implies no anthropization, while values nearing 1 suggest almost complete anthropization. To maintain model simplicity, we hypothesize that substantial anthropization of natural landscapes profoundly affects wild animal dynamics by diminishing their habitat's carrying capacity (see also \cite{dumont2025humanwildlifeinteractionstropicalforest} for a similar consideration), which can ultimately trigger animal dispersal. Furthermore, we posit that the direct influence of anthropization on sylvatic hematophagous vectors (e.g., mosquitoes, tsetse flies) primarily induces vector dispersal. More precisely, since our default assumption is that every patch is a suitable habitat for both vectors and animals, we assume that no displacement of either vectors or animals occurs from patch $i$ if it is not anthropized (i.e. $\alpha_{i}=0$). Moreover, we assume that for low values of $\alpha_{i}$, the displacement rates are very low while for anthropization level $\alpha_{i}$ above a given threshold, we may expect large values of displacement rates for both vectors and animals. In other words, the feedback of the anthropization level on the displacement rates is modeled using the sigmoidal-like functions given in \eqref{mAmV}:
\begin{equation}\label{mAmV}
\begin{array}{l}
m_A(\alpha_{i})=\dfrac{\alpha_{i}^\theta}{\alpha_{i}^\theta+c_{A}^\theta},\\
m_V(\alpha_{i})=\dfrac{\alpha_{i}^\theta}{\alpha_{i}^\theta+c_{V}^\theta}
\end{array}
\end{equation}
where $\theta>1$ controls the shape of the functions, $c_{A}$ (resp. $c_{V}$) is the animal-related (resp. vector-related) half-saturation parameter. The vector and animal dispersal parameters are, respectively (see also \cite{Heath2022,Lutambi2013}),
\begin{equation}\label{dispersal-param}
D_V=\dfrac{D_V^*}{\mathcal{A}} \mbox{  and  } D_A=\dfrac{D_A^*}{\mathcal{A}},
\end{equation}
where $\mathcal{A}$ is the area of each patch (i.e. 4~km$^2$), $D_V^*$ and $D_A^*$ are the average distance traveled per unit time (km$^2$\,day$^{-1}$) by vectors and animals, respectively. We further assume that vector displacement decreases with animal density to reflect the vectors' tendency to stay in habitats where they have blood meals available. A parameter $b$ quantifies the strength of this tendency.

We first assume that individuals leaving patch $i$ can move to any other patch $j\neq i$ of the domain, irrespective of its distance from patch $i$; we refer to this as unconstrained global dispersal. This assumption is not realistic, since individuals can in reality only reach distant patches through successive movements via intermediate patches. However, because every pair of patches is then directly connected by dispersal, it makes the model analytically tractable (see Section~\ref{subsection}). In the numerical simulations in Section~\ref{section:numerical-simulations}, we therefore use the more realistic, but less analytically tractable, regimes of random and constrained neighbor dispersal introduced in Section~\ref{sec:neighbor}, in which individuals only move between neighboring patches. The $n$-patch model with unconstrained global dispersal therefore reads:
\begin{equation}\label{n-patches}
	\left\{
	\begin{array}{l}
\dfrac{dA_i}{dt}=A_ig_i(A_i)+\sum\limits_{j=1, j\neq i}^{n}D_{A}m_A(\alpha_j)A_j-\sum\limits_{{j=1, j\neq i}}^{n}D_{A}m_A(\alpha_i)A_i,\\
\dfrac{dV_i}{dt}=V_ih_i(A_i, V_i)+\sum\limits_{j=1, j\neq i}^{n}D_{V}m_V(\alpha_j)\dfrac{1}{1+bA_j}V_j-\sum\limits_{{j=1, j\neq i}}^{n}D_{V}m_V(\alpha_i)\dfrac{1}{1+bA_i}V_i,\\
A_i(0)\geq0,\quad V_i(0)\geq0,\quad i=1,2,...,n
	\end{array}	
	\right.
\end{equation}
where for every $i\in\{1,2,3,...,n\}$, 
\begin{equation}\label{gi}
\begin{array}{l}
g_i(A_i)=r_{A,i}\left(1-\dfrac{A_i}{(1-\alpha_i)K_{A,i}}\right)-\mu_{A,i},\\
h_i(A_i,V_i)=r_{V,i}\dfrac{A_i}{A_i+a_i}\left(1-\dfrac{V_i}{K_{V,i}}\right)-\mu_{V,i}.\\	
\end{array}
\end{equation}
The parameter $r_{A,i}$ (resp. $r_{V,i}$) denotes the wild animals (resp. vectors) intrinsic growth rate, $K_{A,i}$ stands for the wild animal carrying capacity in the absence of anthropization, $K_{V,i}$ represents the vector carrying capacity under high animal abundance, and $\mu_{A,i}$ (resp. $\mu_{V,i}$) denotes the wild animals (resp. vectors) natural mortality rate. Moreover, the hematophagous vectors ($V_i$) need blood meals from wild animals ($A_i$). Indeed, empirical evidence~\cite{Likeufack2025,lord_host-seeking_2017} suggests that a female will need to feed approximately every 2.5 days before giving birth to a larva. This indicates that there is a relationship between feeding success and the birth rate. We model this positive effect of wild animals on vectors with the term
\begin{align*}
\dfrac{A_i}{A_i+a_i}
\end{align*}
 where $a_i$ is the half-saturation constant.

We set $\mathcal{D}=\mathbb{R}^{2n}_+=\{x\in\mathbb{R}^{2n}:x\geq\textbf{0}\}$. Then model \eqref{n-patches} can be written in the form 

\begin{equation}\label{ODE-system}
\frac{dx}{dt}=f(x),
\end{equation}
where $f:\mathbb{R}^{2n}\to\mathbb{R}^{2n}$ represents the right hand side of (\ref{n-patches}). Function $f$ is continuous and continuously differentiable on $\mathbb{R}^{2n}$. Thus, according to \cite[Theorem III.10.VI]{Walter1998}, for any initial condition a unique solution exists, at least locally. The vector field defined by $f$ is tangential on $\partial D$. Therefore, for any initial condition in $\mathcal{D}$ the solution of (\ref{n-patches}) remains in $\mathcal{D}$ for its maximal interval of existence \cite[Theorem III.10.XVI]{Walter1998}.
	
\subsubsection{The model without dispersal}
Assuming that population dispersal is not considered, system \eqref{n-patches} reduces to:
\begin{equation}\label{without-dispersal}
\left\{
\begin{array}{l}
\dfrac{dV_{i}}{dt}= r_{V,i}\dfrac{A_{i}}{A_{i}+a_{i}}V_{i}\left(1-\dfrac{V_{i}}{K_{V,i}}\right)-\mu_{V,i}V_{i},\qquad i=1,2,3,...,n,\\

\dfrac{dA_{i}}{dt}= r_{A,i}A_{i}\left(1-\dfrac{A_{i}}{(1-\alpha_{i})K_{A,i}}\right)-\mu_{A,i}A_{i},\\
\end{array}
\right.
\end{equation}
 Following \cite{Moulay2011}, the set $\Omega_i$ defined by

\begin{align*}
\Omega_i =\left\{(V_{i},A_{i})\in\mathbb{R}\times \mathbb{R} | 0\leq V_{i} \leq K_{V,i}, 0\leq A_{i} \leq (1-\alpha_{i})K_{A,i} \right\}
\end{align*}
is positively invariant and absorbing for system (\ref{without-dispersal}).
In the absence of dispersal, the dynamics on each patch are independent of those on the other patches. Therefore, in the sequel we give qualitative results on a single patch $i$. Let us consider the following thresholds:
\begin{equation}
\label{thresholds-1}
\mathcal{R}_{A,i}=\dfrac{r_{A,i}}{\mu_{A,i}},\quad \mathcal{R}_{V,i}=\dfrac{r_{V,i}}{\mu_{V,i}}\dfrac{(1-\alpha_{i})K_{A,i}\left(1-\dfrac{1}{\mathcal{R}_{A,i}}\right)}{(1-\alpha_{i})K_{A,i}\left(1-\dfrac{1}{\mathcal{R}_{A,i}}\right)+a_{i}}.
\end{equation}

Straightforward computations lead to the following results on a single patch $i$. 
\begin{theorem}\label{theorem1}
	\begin{enumerate}
		\item The point $E_{00}=(0,0)$ is always an equilibrium of system \eqref{without-dispersal}. It is the extinction equilibrium of the patch $i$.
		\item When $\mathcal{R}_{A,i}>1$, then the point
\begin{align*}
E_{0A}=\left(0,(1-\alpha_{i})K_{A,i}\left(1-\dfrac{1}{\mathcal{R}_{A,i}}\right)\right)
\end{align*}
 is an equilibrium of system \eqref{without-dispersal}. It is the vector-free equilibrium of patch $i$.
		\item Assume that $\mathcal{R}_{A,i}>1$ and $\mathcal{R}_{V,i}>1$. Then, the point
\begin{align*}
E_{VA}=\left(K_{V,i}\left(1-\dfrac{1}{\mathcal{R}_{V,i}}\right),(1-\alpha_{i})K_{A,i}\left(1-\dfrac{1}{\mathcal{R}_{A,i}}\right)\right)
\end{align*}
 is the coexistence equilibrium of system \eqref{without-dispersal}.    
	\end{enumerate}
\end{theorem}

\begin{proof}
	See \cite{Happi2026}.
\end{proof}

\comment{
By computing the jacobian matrices of system \eqref{without-dispersal}, we obtain that the following results are valid.
\begin{theorem}
	\label{theorem-local-stability1p}	
	\begin{enumerate}
		\item The equilibrium $E_{00}$ is locally asymptotically stable (LAS) whenever $\mathcal{R}_{A,i}<1$. It is unstable when $\mathcal{R}_{A,i}>1$.
		\item Assume that the equilibrium $E_{0A}$ exists, that is
		$\mathcal{R}_{A,i}>1$. Then $E_{0A}$ is LAS when $\mathcal{R}_{V,i}<1$ and unstable if $\mathcal{R}_{V,i}>1$.
		\item The equilibrium $E_{VA}$ is LAS whenever it exists, that is when $\mathcal{R}_{A,i}>1$ and $\mathcal{R}_{V,i}>1$.
	\end{enumerate}    
\end{theorem}

\begin{proof}
	See Appendix \ref{AppendixB}.
\end{proof}
}
Using standard comparison arguments (see for instance \cite{Tewa2013} and references therein), we obtain global stability results. 
\begin{theorem}
	\label{theorem-global-stability}
	\begin{enumerate}
		\item The equilibrium $E_{00}$ is globally asymptotically stable (GAS) whenever $\mathcal{R}_{A,i}\leq1$. 
		\item Assume that the equilibrium $E_{0A}$ exists, that is
		$\mathcal{R}_{A,i}>1$. Then $E_{0A}$ is GAS when $\mathcal{R}_{V,i}\leq1$.
		\item The equilibrium $E_{VA}$ is GAS whenever it exists, that is when $\mathcal{R}_{A,i}>1$ and $\mathcal{R}_{V,i}>1$.
	\end{enumerate}    
\end{theorem}

\begin{proof}
	See \cite{Happi2026}. 
\end{proof}

These results are summarized in Table \ref{tab:stabilite1p}.

\begin{table}[H]
	\centering
	\begin{tabular}{|c|c|c|c|c|}
		\hline
		$\mathcal{R}_{A,i}$ & $\mathcal{R}_{V,i}$ & $E_{00}$ & $E_{0A}$ & $E_{VA}$\\
		\hline
		$\leq1$ & & GAS &&\\
		\hline
		$>1$ & $\leq1$ & Unstable & GAS & \\
		\cline{2-5}
		& $>1$ & Unstable &Unstable & GAS \\
		\hline
	\end{tabular}
	\caption{Summary of the stability results for the equilibria of system \eqref{without-dispersal}.}
	\label{tab:stabilite1p}
\end{table}

\subsubsection{The model with dispersal}\label{subsection}	

The equations describing animal dynamics in \eqref{n-patches} represent an $n$-patch single-species model that has already been the subject of several works, see e.g. \cite{Bliman2025-uf,Elbetch2023-ep,Elbetch2021-ru}. Some meaningful background concepts and results dealing with matrices are recalled in Appendix \ref{AppendixD}. For ease of notation, we denote below by $0_{\mathbb{R}^n}$ the $n$-dimensional zero vector. Let us set $A=(A_i)$, $V=(V_i)$, $i=1,2,3,...,n$. We also define the matrices $d^A=(d_{ij}^A)$ and $d^V(A)=(d_{ij}^V)$ where
\begin{equation}\label{mat-dA}
	d_{ij}^A=\left\{\begin{array}{ll}
	D_{A}m_A(\alpha_j) & \mbox{ if } i\neq j,\\
	-\sum\limits_{k=1, k\neq i}^{n}D_Am_A(\alpha_i) & \mbox{ if } i= j,\\
	\end{array}
	\right.
\end{equation}

and 

\begin{equation}\label{mat-dV}
d_{ij}^V=\left\{\begin{array}{ll}
D_{V}m_V(\alpha_j)\dfrac{1}{1+bA_j} & \mbox{ if } i\neq j,\\
-\sum\limits_{k=1, k\neq i}^{n}D_Vm_V(\alpha_i)\dfrac{1}{1+bA_i} & \mbox{ if } i= j.\\
\end{array}
\right.
\end{equation}

System \eqref{n-patches} in compact form reads as
\begin{equation}\label{n-patches-compact}
\left\{
\begin{array}{l}
\dfrac{dA}{dt}=\left(d^A+\operatorname{diag}(g_i(A_i))\right)A,\\
\dfrac{dV}{dt}=\left(d^V(A)+\operatorname{diag}(h_i(A_i,V_i))\right)V.
\end{array}	
\right.
\end{equation}
We define
\begin{align*}
M_A=d^A+\operatorname{diag}(g_i(0_{\mathbb{R}^n})).
\end{align*}
 For a square matrix $M$, we denote by $s$ its stability modulus defined by 
\begin{align*}
s(M)=\max \{\operatorname{Re}(\lambda): \lambda\in\sigma(M)\}
\end{align*}
where $\sigma(M)$ is the set of all eigenvalues of $M$, see also Definition \ref{def1} in Appendix \ref{AppendixD}. Equilibria of system \eqref{n-patches-compact} are summarized in Theorem \ref{theorem:equilibria} and their stability properties are presented in Theorem \ref{theorem:LAS}.

\begin{theorem}\label{theorem:equilibria} 
\begin{enumerate}
	\item If $s(M_A)\leq 0$, then the point $E_0=(0_{\mathbb{R}^n},0_{\mathbb{R}^n})$ is the unique equilibrium of system \eqref{n-patches-compact}.
	\item Assume that $s(M_A)> 0$, then the uncoupled equation \eqref{n-patches-compact}$_1$ admits a unique positive equilibrium $A^*$. In addition, we set
\begin{align*}
M_V=d^V(A^*)+\operatorname{diag}(h_i(A^*,0_{\mathbb{R}^n})).
\end{align*}
	\begin{enumerate}
		\item If $s(M_V)\leq 0$, then system \eqref{n-patches-compact} admits two equilibria, $E_0=(0_{\mathbb{R}^n},0_{\mathbb{R}^n})$ and $E_1=(A^*,0_{\mathbb{R}^n})$ where $\min(A^*)>0$. 
		\item If $s(M_V)> 0$, then system \eqref{n-patches-compact} admits two equilibria, $E_0=(0_{\mathbb{R}^n},0_{\mathbb{R}^n})$ and $E_2=(A^*,V^*)$ where $\min(A^*)>0$ and $\min(V^*)>0$. 
	\end{enumerate}	
\end{enumerate}
\end{theorem}

\begin{proof}
System \eqref{n-patches-compact} is a triangular system. Since assumptions (H1)--(H3) of Appendix \ref{AppendixD} hold, the uncoupled equation \eqref{n-patches-compact}$_1$ is such that, if $s(M_A)\leq0$ then $0_{\mathbb{R}^n}$ is its unique equilibrium and if $s(M_A)>0$ system \eqref{n-patches-compact}$_1$ admits two equilibria $0_{\mathbb{R}^n}$ and $A^*$ (see Theorem \ref{theorem:n-patches-single-species} in Appendix \ref{AppendixD}). By using a similar argument and by assuming that $s(M_A)>0$ we obtain that if  $s(M_V)\leq0$ then $0_{\mathbb{R}^n}$ is the unique equilibrium of system \eqref{n-patches-compact}$_2$ where $A:=A^*$. Moreover, for $s(M_V)>0$, then system \eqref{n-patches-compact}$_2$, with $A:=A^*$, has two equilibria: $0_{\mathbb{R}^n}$ and $\min(V^*)>0$. Note that system \eqref{n-patches-compact}$_2$, with $A:=0_{\mathbb{R}^n}$, has a unique equilibrium which is $0_{\mathbb{R}^n}$. This completes the proof of the Theorem.
\end{proof}

\begin{theorem}\label{theorem:LAS}
	\begin{enumerate}
		\item If $s(M_A)\leq 0$ then the point $E_0=(0_{\mathbb{R}^n},0_{\mathbb{R}^n})$ is locally asymptotically stable (LAS).
		\item If $s(M_A)> 0$ and $s(M_V)\leq 0$, then $E_0=(0_{\mathbb{R}^n},0_{\mathbb{R}^n})$ is unstable and $E_1=(A^*,0_{\mathbb{R}^n})$ is LAS.
		\item If $s(M_A)> 0$ and $s(M_V)> 0$, then $E_0=(0_{\mathbb{R}^n},0_{\mathbb{R}^n})$ is unstable and $E_2=(A^*,V^*)$ is LAS.	
	\end{enumerate}
\end{theorem}

\begin{proof}
	System \eqref{n-patches-compact} is a triangular system. Since assumptions (H1)--(H3) of Appendix \ref{AppendixD} hold, then the proof follows from a direct application of Theorem \ref{theorem:n-patches-single-species} and Theorem \ref{Vidyasagar-Theorem}, recalled in Appendix \ref{AppendixD}.
\end{proof}

	
\subsection{The $n$-patch model with random and constrained neighbor dispersal}\label{sec:neighbor}

In this section, we adapt our model by restricting dispersal to neighboring patches. Here, $n_i$ denotes the set consisting of patch $i$ and the eight patches (unless on domain boundary) that share an edge or a corner with it. We denote by $q_y$ the proportion of the animals leaving patch $i$ that move into patch $y\in n_i\setminus\{i\}$, and we use $q_y$ to distinguish two dispersal regimes. Under random neighbor dispersal, animals choose any neighboring patch with equal probability, that is
\begin{equation}\label{q-random}
q_y=\dfrac{1}{|n_i\setminus\{i\}|},
\end{equation}
where $|n_i\setminus\{i\}|$ is the number of neighbors of patch $i$. Under constrained neighbor dispersal, we instead assume that animals condition their migration targets according to the anthropization levels of all neighboring patches. While we leave the outward migration rate from a patch $i$ unchanged, we assume that the destination of animal migration within the set of neighboring patches $n_i$ is positively correlated with the ``quality index'' of each $y\in n_{i}\setminus\{i\}$. Specifically, for $y\in n_{i}\setminus\{i\}$, its quality index is defined by
\begin{equation}\label{quality-index}
q_y=\dfrac{1-\alpha_{y}^\gamma}{\sum\limits_{l\in n_i, l\neq i}(1-\alpha_l^\gamma)}, \qquad \gamma>1.
\end{equation}
Hence, under constrained neighbor dispersal, $q_y$ is a decreasing function of $\alpha_y$ and $\gamma$ shapes the function $q_y$. In both regimes, $q_y$ also depends on the patch of origin $i$ (through the number of neighbors of $i$ or through the normalization in the denominator); we omit this dependence for ease of notation. We further assume that vectors cannot make such constrained movement decisions, i.e. vectors always follow random neighbor dispersal.

The movement rates from patch $i$ into a neighboring patch $y\in n_{i}\setminus \{i\}$ 
 are defined by:
\begin{align}
\beta^V_{[i\rightarrow y]}&= D_V,\\
\beta^A_{[i\rightarrow y]}&= |n_i\setminus\{i\}|\, D_A q_y.
\end{align}

Therefore, the total displacement rates out of the patch  $i$ are:
\begin{align}
Out_i^V&:=\left(\sum\limits_{y\in n_{i}\setminus\{i\}}\beta^V_{[i\rightarrow y]}\right)m_V(\alpha_{i})\dfrac{1}{1+bA_{i}}, \\
Out_i^A&:=\left(\sum\limits_{y\in n_{i}\setminus\{i\}}\beta^A_{[i\rightarrow y]}\right)m_A(\alpha_{i}).
\end{align}
Since the $q_y$ sum to one, $Out_i^A=|n_i\setminus\{i\}|D_Am_A(\alpha_{i})$ under both dispersal regimes, which therefore differ only in the destination of emigrating animals. For random neighbor dispersal, \eqref{q-random} gives $\beta^A_{[i\rightarrow y]}=D_A$, as for vectors.

This gives

\begin{equation}\label{with-dispersal-2}
\left\{
\begin{array}{lcl}
\dfrac{dV_{i}}{dt} &=& r_{V,i}\dfrac{A_{i}}{A_{i}+a_{i}}V_{i}\left(1-\dfrac{V_{i}}{K_{V,i}}\right)-\mu_{V,i}V_{i}\\

&-&\left(\sum\limits_{y\in n_{i}\setminus\{i\}}\beta^V_{[i\rightarrow y]}\right)m_V(\alpha_{i})\dfrac{1}{1+bA_{i}}V_{i}+
\left(\sum\limits_{y\in n_{i}\setminus\{i\}}\beta^V_{[y\rightarrow i]}m_V(\alpha_{y})\dfrac{1}{1+bA_{y}}V_{y}\right),\\

\dfrac{dA_{i}}{dt}&=& r_{A,i}A_{i}\left(1-\dfrac{A_{i}}{(1-\alpha_{i})K_{A,i}}\right)-\mu_{A,i}A_{i}\\
&-&\left(\sum\limits_{y\in n_{i}\setminus\{i\}}\beta^A_{[i\rightarrow y]}\right)m_A(\alpha_{i})A_{i}+
\left(\sum\limits_{y\in n_{i}\setminus\{i\}}\beta^A_{[y\rightarrow i]}m_A(\alpha_{y})A_{y}\right),\\
\end{array}
\right.
\end{equation}
where $1\leq i\leq n$.

For both random and constrained neighbor dispersal, we set again $A=(A_i)$, $V=(V_i)$, $i=1,2,3,...,n$. We also define the matrices  $d^{A}_{\mathrm{neigh}} = (d_{iy,\mathrm{neigh}}^{A})$ and $d^{V}_{\mathrm{neigh}}(A) = (d_{iy,\mathrm{neigh}}^{V})$ where
\begin{align*}
d_{iy,\mathrm{neigh}}^{A} = \begin{cases}
-(\sum_{y \in n_i\setminus\{i\}}\beta_{[i \rightarrow y]}^{A})m_A(\alpha_i)  &\text{if} \hspace{1 cm}i=y,\\
\beta_{[y \rightarrow i]}^{A}m_A(\alpha_y)			&\text{if} \hspace{1 cm}y \in n_i\setminus\{i\},\\
0							&\text{if} \hspace{1 cm}i\neq y \hspace{0.1 cm}\textrm{and}\hspace{0.1 cm} y \notin n_i\setminus\{i\},
\end{cases}
\end{align*}
\begin{align*}
d_{iy,\mathrm{neigh}}^{V} = \begin{cases}
-(\sum_{y \in n_i\setminus\{i\}}\beta_{[i \rightarrow y]}^{V})m_V(\alpha_i)\dfrac{1}{1+bA_i}  &\text{if} \hspace{1 cm}i=y,\\
\beta_{[y \rightarrow i]}^{V}m_V(\alpha_y)\dfrac{1}{1+bA_y}			&\text{if} \hspace{1 cm}y \in n_i\setminus\{i\},\\
0							&\text{if} \hspace{1 cm}i\neq y \hspace{0.1 cm}\textrm{and}\hspace{0.1 cm} y \notin n_i\setminus\{i\}.
\end{cases}
\end{align*}

With these notations, the system \eqref{with-dispersal-2} becomes
\begin{equation}\label{system_red}
\begin{cases}
\dfrac{dA}{dt} &= (d^A_{\mathrm{neigh}} + \operatorname{diag}(g_i(A_i)))A,\\
\dfrac{dV}{dt} &= (d^V_{\mathrm{neigh}}(A) + \operatorname{diag}(h_i(A_i, V_i)))V.
\end{cases}
\end{equation}


Let us define
\begin{align*}
M_{A,\mathrm{neigh}} = d^{A}_{\mathrm{neigh}} + \operatorname{diag}(g_i(0_{\mathbb{R}^n})).
\end{align*}
Since  the movements of both vectors and animals present on a current patch $i$ are restricted only to neighboring patches $n_i$, the graph of the patches might not be strongly connected; in other words, the matrix $M_{A,\mathrm{neigh}}$ may be reducible. However, in the case where $M_{A,\mathrm{neigh}}$ is not irreducible, then by renumbering
the patches, it can be given a block triangular structure \cite{Auger2008-lc, berman_nonnegative_1994}. Indeed, based on \cite{berman_nonnegative_1994}, when $M_{A,\mathrm{neigh}}$ is reducible,  there exists a suitable permutation matrix $P$ such that $M_{A,\mathrm{neigh}}$ can be reduced to the triangular block form,
\begin{equation}
PM_{A,\mathrm{neigh}}P^T =
\begin{pmatrix}
M_{11} & 0 & \cdots & \cdots & 0\\
M_{21} & M_{22} & 0 & \cdots & 0\\
\vdots & \vdots & \vdots &  \ddots &  \vdots\\
M_{s1} & M_{s2} & \cdots & & M_{ss}\\
\end{pmatrix}
\end{equation}
where the diagonal blocks $M_{ii}$ are irreducible.
Suppose that the blocks $M_{ii}$ correspond to the traps, which are groups of compartments for which there are no transfers out of the group. Therefore, in the long-term, all the material will be transferred into the system's traps. Hence, the diagonal blocks will behave as irreducible classes. To study the asymptotic properties of the system, it is then enough to limit the study to the traps. All this shows that $M_{A,\mathrm{neigh}}$ can be assumed to be irreducible, see also \cite{Auger2008-lc}. Therefore, results in Theorems \ref{theorem:equilibria}-\ref{theorem:LAS} are still valid with $M_{A,\mathrm{neigh}}$ and
\begin{align*}
M_{V,\mathrm{neigh}}=d^V(A^*)_{\mathrm{neigh}}+\operatorname{diag}(h_i(A^*,0_{\mathbb{R}^n}))
\end{align*}
when referring to $M_{A}$ and $M_{V}$, respectively.

\section{Numerical simulations: a focus on \textit{Glossina} spp.}\label{section:numerical-simulations}
In this section, we focus on \textit{Glossina}, commonly referred to as tsetse flies. Tsetse flies (\textit{Glossina} spp.) are hematophagous insects endemic to sub-Saharan Africa and are the primary vectors of \textit{Trypanosoma} parasites. These parasites are responsible for Human African Trypanosomiasis (HAT), commonly called sleeping sickness, and Animal African Trypanosomiasis (AAT), also known as nagana \cite{gooding_tsetse_2005}. HAT is a neglected tropical disease that, if left untreated, leads to severe neurological disorders and eventually death, while AAT remains one of the greatest constraints to livestock production across the continent \cite{Ilboudo2022-jt}. The dual burden of these diseases imposes serious challenges on public health, food security, and rural development. 

Hematophagy, or blood-feeding, is essential for the survival of \textit{Glossina}. They feed on a wide range of vertebrates, including humans, other mammals (e.g.\ bushbuck, rhinoceros, equines, bovids and porcupines) and reptiles \cite{laveissiere_glossines_nodate-1}. 

Tsetse flies fly only for short periods each day \cite{maudlin_tsetse_2004,laveissiere_glossines_nodate-1}, about 20 to 25 minutes in total \cite{jackson_biology_1949}. However, their dispersal varies with species, climatic conditions and habitat change \cite{laveissiere_glossines_nodate-1}. For example, it is shown in \cite{laveissiere_glossines_nodate-1} that individual \textit{Glossina palpalis} can travel up to 17~km in just three days.    Gibson et al. \cite{Gibson1991} found that tsetse flies are capable of flying at around 20~km/h and are active for about 30 minutes daily.
This mobility explains why tsetse-cleared areas are rapidly reinvaded. For example, aerial spraying that eliminated almost all flies in a 4~km$^2$ area in Zimbabwe in the 1980s \cite{Vale1984} proved temporary, with populations returning to normal within a day due to immigration from neighboring areas. Thus, studying or controlling tsetse over small areas is difficult. Tsetse flies are particularly sensitive to olfactory stimuli, using their antennae to detect carbon dioxide (CO$_2$) and other volatile compounds emitted by potential hosts. These olfactory cues are crucial for host detection, allowing the flies to home in on animals from several meters away \cite{jackson_biology_1949}.

Tsetse flies primarily inhabit savannas, forests, and riverine environments where humidity and shade are sufficient for survival. Their population in a given locality depends on climatic conditions, human activity and vegetation cover. In particular, humidity strongly affects the distribution of tsetse \cite{akinseye_population_2024}, and tsetse flies prefer to move toward areas with high humidity or less anthropization. More specifically, tsetse flies prefer to live on tree trunks or where vegetation cover is high.

Following fertilization, the egg takes approximately 50 to 60 hours to develop into a larva. The larva then undergoes development over $6$ to $8$ days before being deposited by the female. Immediately after larviposition, the third instar larva burrows into the soil and pupates. The puparial stage lasts about 28 days for males and 26 days for females. After emergence, the fly undergoes a maturation period of approximately 3 days for females and 7 days for males before becoming reproductively active \cite{Likeufack2025}. 

The effective birth rate ($r_V$) is defined as the product of the proportion of viable larvae, the probability of surviving the larval and pupal stages, and the ratio of the average larva production per female lifetime to the average female lifespan. The survival probabilities for the larval and pupal stages are $p\approx0.95$ and $q\approx0.75$, respectively \cite{Likeufack2025}. According to \cite{hargrove_reproductive_1994, Likeufack2025}, the abortion proportion is assumed to be $0.025$. Therefore the proportion of viable larvae will be $1-0.025 = 0.975$. 



Based on \cite{jackson_biology_1949,leak_tsetse_nodate}, the lifespan of an adult female (resp. male) is around 105 days (resp. 28 days). Gooding and Krafsur \cite{gooding_tsetse_2005} estimated that a female produces about 7 larvae in her lifetime. Therefore, we have $r_V = \dfrac{7}{105} \times p \times q\times(1-0.025) = 0.0463$. We assumed the mortality rate to be the reciprocal of the average female lifespan, $\mu_V= \dfrac{1}{105}\approx 0.00952$. Re-interpreting results of \cite{Cecilia2021}, we deduce that patch carrying capacity for tsetse is approximately $K_V=5000$ tsetse individuals. Based on \cite{Duda2017-er,dumont2025humanwildlifeinteractionstropicalforest,Robinson1986-jb}, we assume that the wild animals' (mostly forest mammals) intrinsic growth rate $r_A\in[0.0012, 0.0023]$ and their carrying capacity $K_A$ to be around 1000 individuals \cite{dumont2025humanwildlifeinteractionstropicalforest}.

Therefore, for $k=1,2,...,n$, we assume that
\begin{align*}
r_V(k) &= 0.0463+0.01 \times rand(),\quad (\text{in day}^{-1})\\
\mu_V(k) &= 0.00952 + 0.0001 \times rand(), \quad (\text{in day}^{-1})\\
K_V(k) &= 50 + 5 \times rand(), \quad (\text{in individuals} \times 100)\\
r_A(k) &= 0.0015+0.0005 \times rand(), \quad (\text{in day}^{-1})\\
\mu_A(k) &= 0.0002 + 0.00001 \times rand(), \quad (\text{in day}^{-1})\\
K_A(k) &= 10 + 1 \times rand(), \quad (\text{in individuals} \times 100)\\
a(k) &= 3+ 2\times rand(), \quad (\text{in individuals} \times 100)
\end{align*}
where $rand()\in[0,1)$ returns a uniformly distributed random number between $0$ and $1$ for each patch in the domain. For all numerical simulations that follow, we used the same parameter values in each simulation. That is, the random draw of parameter values causes differences between individual patches in the spatial domain, but no differences between separate realizations of the model. Moreover, we assumed $c_A = 0.03$, $c_V = 0.04$, $\theta=2$, $b=0.01$, $D_V = 0.01$, $D_A = 0.001$, and $\gamma = 2$ in each patch without variation between patches.

\subsection{The case of random neighbor dispersal} \label{sec: sim unconstrained}
We first simulated the model with random neighbor dispersal of vectors and animals, i.e. system \eqref{system_red} with $q_y$ given by \eqref{q-random}. In the interest of studying how a newly introduced vector population and its animal host may spread throughout the domain, we initialized each simulation by setting both vector and animal densities to their respective carrying capacities in a single patch only, with the population being absent from any other patch initially.

\begin{figure}
\centering
\includegraphics[height=0.9\textheight]{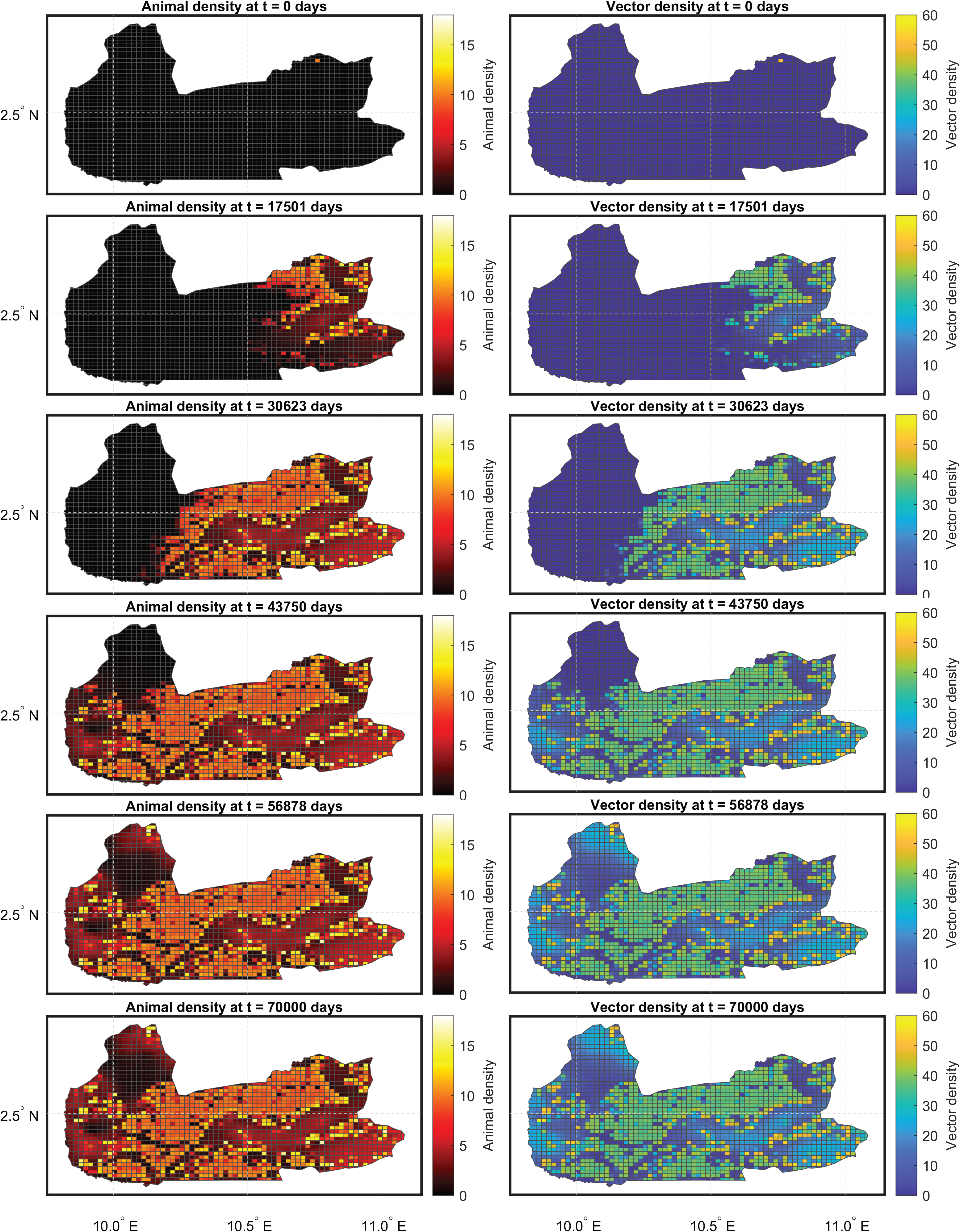}
\caption{Example simulation in the case of random neighbor dispersal. A simulation of system \eqref{system_red} (model with random neighbor dispersal) is shown. The simulation is initialized with the population being absent from all patches except a single patch in the northeast ($2.759^\circ$ N, $10.760^\circ$ E) of the domain. The population gradually invades the whole domain and converges to a steady state.}\label{fig: example sim unconstrained}
\end{figure}

Initializing the simulation with the population being absent from the domain except for a single patch in the northeast ($2.759^\circ$ N, $10.760^\circ$ E), we observed a gradual invasion process of both animals and vectors throughout the whole domain. The simulation eventually converges to a stable configuration (\cref{fig: example sim unconstrained}). This steady state did not depend on the initial condition in our simulations (\cref{fig: appendix ic tests}), consistent with Theorems \ref{theorem:equilibria}--\ref{theorem:LAS}.

\begin{figure}
    \centering
    \includegraphics[width=0.99\linewidth]{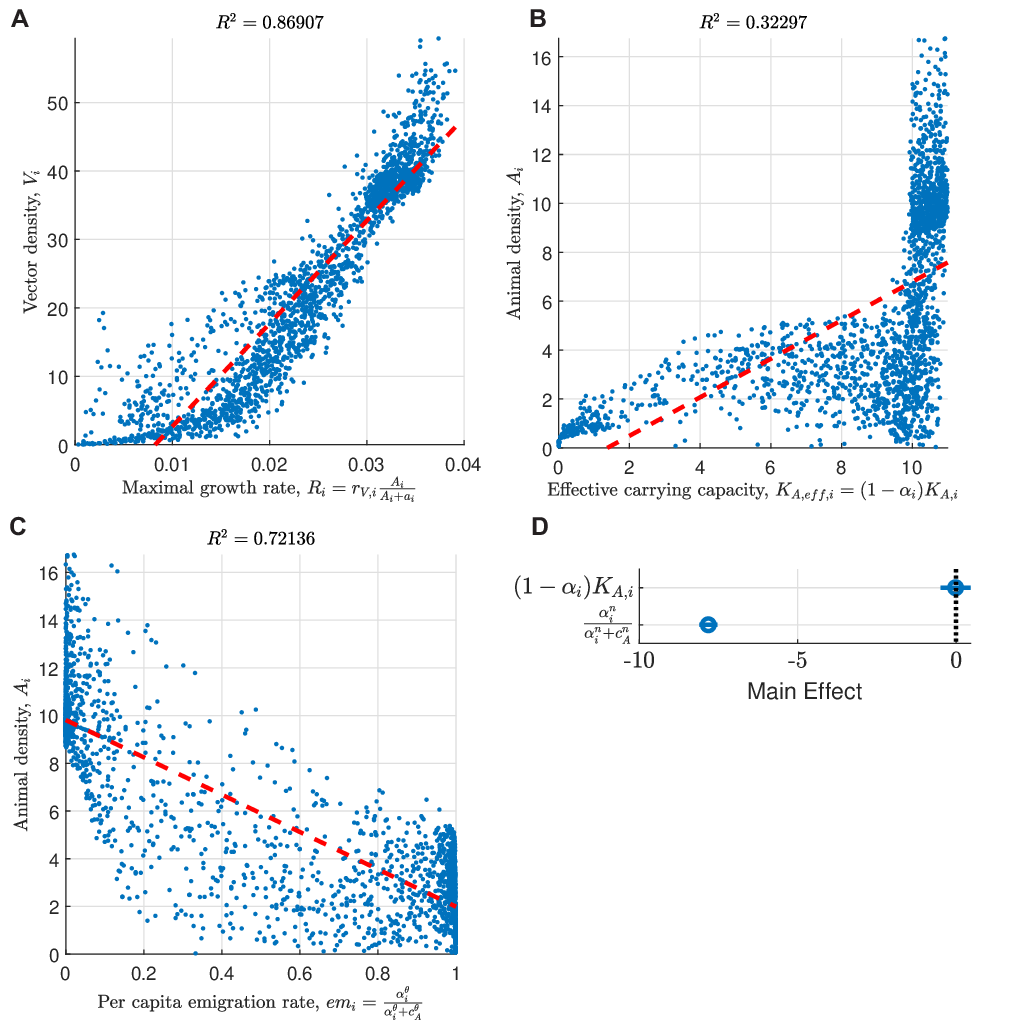}
    \caption{Statistical analysis of steady state densities in the case of random neighbor dispersal (model \eqref{system_red} with \eqref{q-random}). \textbf{A}: The steady state vector density against the vector maximal growth rate $R_i=r_{V,i}\frac{A_i}{A_i+a_i}$ is shown. The line of best fit (red dashed) is $V_i \approx -12 + 1504R_i$. \textbf{B}: The steady state animal density against the animal effective carrying capacity $K_{A,\operatorname{eff},i} = (1-\alpha_i)K_{A,i}$ is shown. The line of best fit is $A_i \approx -1.09 + 0.79K_{A,\operatorname{eff},i}$. \textbf{C}: The steady state animal density against the animal per-capita emigration rate $\operatorname{em}_i := \frac{\alpha_i^\theta}{\alpha_i^\theta + c_A^\theta}$ is shown. The line of best fit is $A_i \approx 9.81 - 7.82\operatorname{em}_i$. \textbf{D}: The main effect sizes of animal effective carrying capacity and animal per-capita emigration rate on animal steady state density are shown, obtained  via multiple linear regression.}
    \label{fig: ss characterisation unconstrained}
\end{figure}

We characterized the steady state solution by recording steady state densities in each patch. First, we established a correlation between the vector and animal densities at steady state. Given that the maximal growth rate of the vector population in patch $i$ (i.e. its per-capita birth rate in the absence of density dependence; mortality is not subtracted) is $R_i:=r_{V,i}\frac{A_i}{A_i+a_i}$, we performed a linear regression between $V_i$ and $R_i$. This showed that 
\begin{equation}
    \label{fit-1}
    V_i \approx -12 + 1504R_i, \quad R^2 = 0.87
\end{equation}
provided the best fit (\cref{fig: ss characterisation unconstrained}A). It is unsurprising to observe a linear relation with negative intercept due to the order of magnitude difference in the mobility coefficients. Vectors disperse faster than animals and thus nearby patches with high vector populations act as stronger sources than nearby patches with high animal densities. This enables vectors to persist despite low animal populations due to migration. 

We next examined how the animal steady state density (and thus also the vector steady state density) depended on the anthropization in the patch. We first compared the animal steady state density with its effective carrying capacity $K_{A,\operatorname{eff},i} = (1-\alpha_i)K_{A,i}$. Linear regression showed that animal steady state density was increasing with $K_{A,\operatorname{eff},i}$ (line of best fit is $A_i \approx -1.09 + 0.79K_{A,\operatorname{eff},i}$), but showed only a weak linear association ($R^2 \approx 0.32$) (\cref{fig: ss characterisation unconstrained}B). This was particularly pronounced for high effective carrying capacity at which there was large variation in the animal steady state density. We thus hypothesized that migration had a larger effect on the steady state densities. To test this, we also compared the animal steady state densities with the per-capita emigration rate, which is proportional to $\operatorname{em}_i := m_A(\alpha_i) = \frac{\alpha_i^\theta}{\alpha_i^\theta + c_A^\theta}$. Linear regression showed a negative dependence with 
\begin{align*}
A_i \approx 9.81 - 7.82\operatorname{em}_i, \quad R^2 \approx 0.72
\end{align*}
being the line of best fit (\cref{fig: ss characterisation unconstrained}C). For better comparison, we also performed a multiple linear regression. This showed that the animal effective carrying capacity had no significant effect on animal steady state density ($p>0.05$) (\cref{fig: ss characterisation unconstrained}D). Using multiple linear regression despite the dependence of both variables on the anthropization level $\alpha$ was valid in this case, as the variance inflation factor was $1.81$, which was below even the most conservative upper limits given in the literature for indicating a significant impact of collinearity ($2.5$ \cite{Johnston2018}; $3$ \cite{Hair2021}).

\subsection{The case of constrained neighbor dispersal}\label{sec: sim constrained}

\begin{figure}
\centering
\includegraphics[height=0.9\textheight]{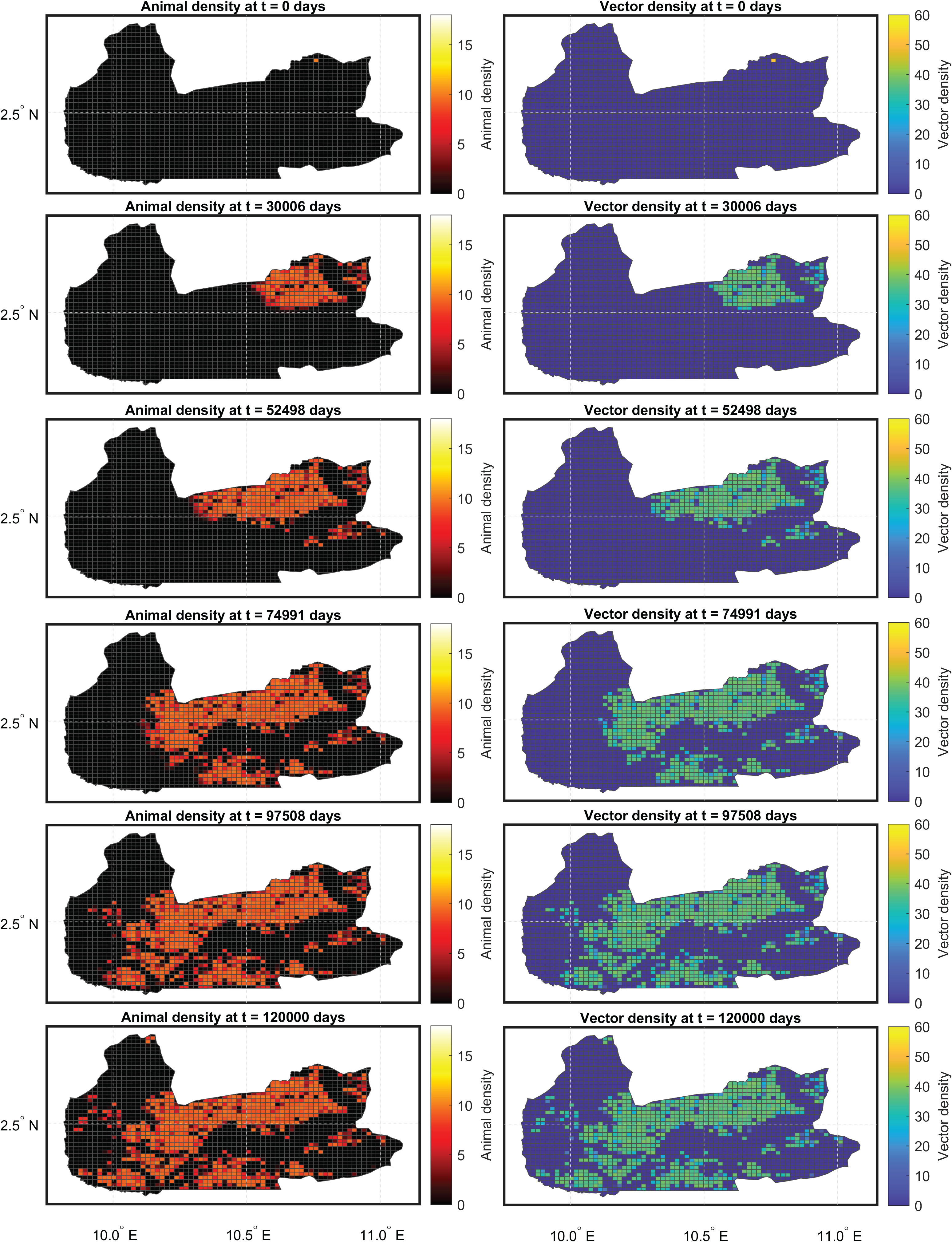}
\caption{Example simulation in the case of constrained neighbor dispersal. A simulation of \eqref{system_red} (model with constrained neighbor dispersal) is shown. The simulation is initialized with the population being absent from all patches except a single patch in the northeast ($2.759^\circ$ N, $10.760^\circ$ E) of the domain. The population gradually spreads through the domain and converges to a steady state in which many patches retain very low densities.}\label{fig: example sim constrained}
\end{figure}

We next turned our attention to the model with constrained neighbor dispersal, i.e. \eqref{system_red} with $q_y$ given by \eqref{quality-index}. For this model, we repeated our numerical simulation regime described above to compare results between the two modes of dispersal. Also with constrained neighbor dispersal, we observed a gradual invasion of both the animal and vector populations throughout the domain (\cref{fig: example sim constrained}). As before, the population eventually converged to a steady state independent of the initial condition (\cref{fig: appendix ic tests constrained}). In contrast to the case of random neighbor dispersal, invasion speed was lower and many patches remained at very low population densities. 

\begin{figure}
    \centering
    \includegraphics[width=0.99\linewidth]{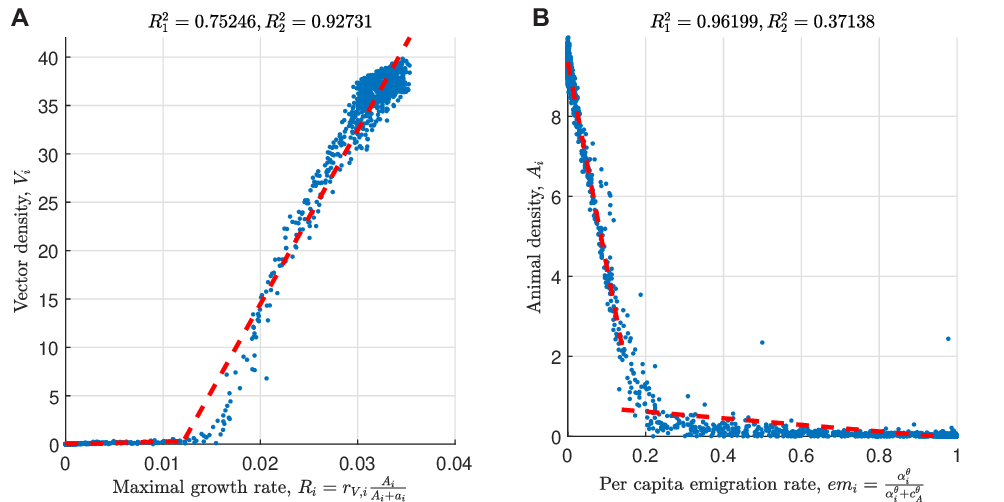}
    \caption{Statistical analysis of steady state densities for constrained neighbor dispersal (model \eqref{system_red} with \eqref{quality-index}). \textbf{A}: The steady state vector density against the vector maximal growth rate $R_i=r_{V,i}\frac{A_i}{A_i+a_i}$ is shown. For expressions of the lines of best fit obtained through piecewise linear regression see \eqref{eq: animal vs vector piecewise linear}. \textbf{B}: The steady state animal density against the animal per-capita emigration rate $\operatorname{em}_i := \frac{\alpha_i^\theta}{\alpha_i^\theta + c_A^\theta}$ is shown. For expressions of the lines of best fit obtained through piecewise linear regression see \eqref{eq: animal vs emigration piecewise linear}.}
    \label{fig: ss characterisation constrained}
\end{figure}

We again first investigated how vector and animal steady state densities depended on each other. Vector density and vector maximal growth rate $R_i=r_{V,i}\frac{A_i}{A_i+a_i}$ showed a remarkably clear piecewise linear relationship (\cref{fig: ss characterisation constrained}A). Piecewise linear regression showed that 
\begin{equation} \label{eq: animal vs vector piecewise linear}
    V_i \approx \begin{cases}
    25 R_i \quad &R^2 \approx 0.75, \quad \text{if} \quad R_i<0.012, \\
    -21 + 1795 R_i \quad &R^2 \approx 0.93, \quad \text{if} \quad R_i>0.012,
\end{cases}
\end{equation}
Given the importance of the per-capita emigration rate $\operatorname{em}_i = \frac{\alpha_i^\theta}{\alpha_i^\theta + c_A^\theta}$ in the case of random neighbor dispersal, we next examined the relation between animal steady state density and $\operatorname{em}_i$. This also showed a remarkably clear piecewise linear relationship (\cref{fig: ss characterisation constrained}B) with 
\begin{equation}\label{eq: animal vs emigration piecewise linear}
A_i \approx \begin{cases}
    9.4 - 51 \operatorname{em}_i \quad &R^2 \approx 0.96, \quad \text{if} \quad \operatorname{em}_i<0.14, \\
    0.78 - 0.82 \operatorname{em}_i \quad &R^2 \approx 0.37, \quad \text{if} \quad \operatorname{em}_i>0.14,
\end{cases}
\end{equation}
Given the very strong correlation between far-from-zero animal steady state density values and per-capita emigration rate, we did not perform any further statistical tests.

\subsection{Some anthropization scenarios}
\begin{figure}
    \centering
    \includegraphics[width=0.99\linewidth]{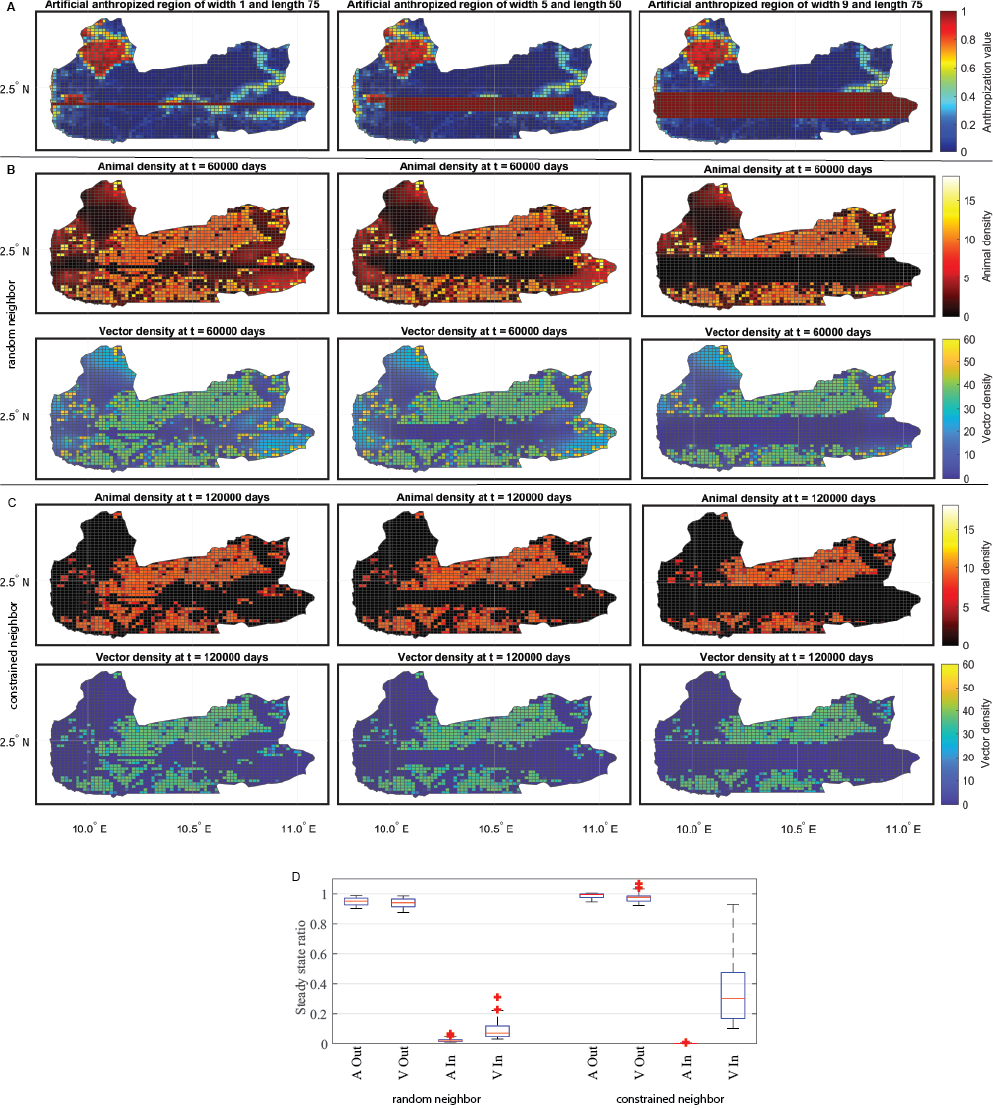}
    \caption{Example simulations in a domain with an artificially anthropized region. \textbf{A}: The different domains used in the example simulations are shown. \textbf{B}: Steady state distributions (at $t=60{,}000$ days) for each simulation (corresponding to the domain shown in the same column) are shown for simulations of the model \eqref{system_red} with random neighbor dispersal. \textbf{C}: Steady state distributions (at $t=120{,}000$ days) for each simulation (corresponding to the domain shown in the same column) are shown for simulations of the model \eqref{system_red} with constrained neighbor dispersal. \textbf{D:} Boxplots of animal and vector steady state ratios both outside and inside the artificially anthropized regions. The left column shows data for the case of random neighbor dispersal, the right column shows data for the case of constrained neighbor dispersal. } \label{fig: anthropization examples}
\end{figure}

We next considered some artificial anthropization scenarios. For this, we changed the landscape used in our simulation by introducing areas of near maximal anthropization ($\alpha = 0.999$ is used as an approximation to $\alpha = 1$ to avoid singularities) in the domain. We used rectangular anthropized regions centered at $2.39^\circ$ N, $10.427^\circ$ E with widths (north-south) ranging from $1$ to $9$ grid patches (increment $1$) and lengths (west-east) ranging from $5$ to $75$ grid patches (increment $5$). Some example regions are shown in \cref{fig: anthropization examples}A. As before, we simulated model \eqref{system_red} with both random neighbor dispersal and constrained neighbor dispersal to steady state. Some example steady states are shown in \cref{fig: anthropization examples}B and C, respectively.

We compared the steady state densities obtained through these simulations on domains with artificially anthropized regions with the steady states of our baseline simulations performed in \cref{sec: sim unconstrained,sec: sim constrained}. For this, we separated the domain patches into those that featured an increased anthropization level $\alpha_i$ compared with the baseline simulations and those that had identical parameter values. For each simulation, we calculated the ratio of steady state densities between the new simulations and the baseline case for each patch and recorded the mean ratio within each of the two groups of patches. This showed that in both the random neighbor dispersal and constrained neighbor dispersal cases, patches outside the artificially anthropized region were only marginally affected, while patches within the anthropized region experienced a significant decrease in steady state densities of both populations (\cref{fig: anthropization examples}D). In the case of random neighbor dispersal, all simulations resulted in patches outside the anthropized region experiencing a slight decrease in steady state densities for both vectors and animals on average, likely because the anthropized region, in which the effective carrying capacity is close to zero, acts as a population sink: it no longer supplies migrants to the surrounding patches, while individuals that move into it are lost. Most simulations in the constrained neighbor dispersal case showed a similar trend, but some simulations led to a slight increase in vector and animal steady state densities when compared to the baseline simulations. Overall, the impact on animal densities was more pronounced than on vector densities because only animal movement was constrained. However, owing to the dependence of vectors on animals, vector steady state densities inside the anthropized region were also strongly reduced.

\begin{figure}
    \centering
    \includegraphics[width=0.75\linewidth]{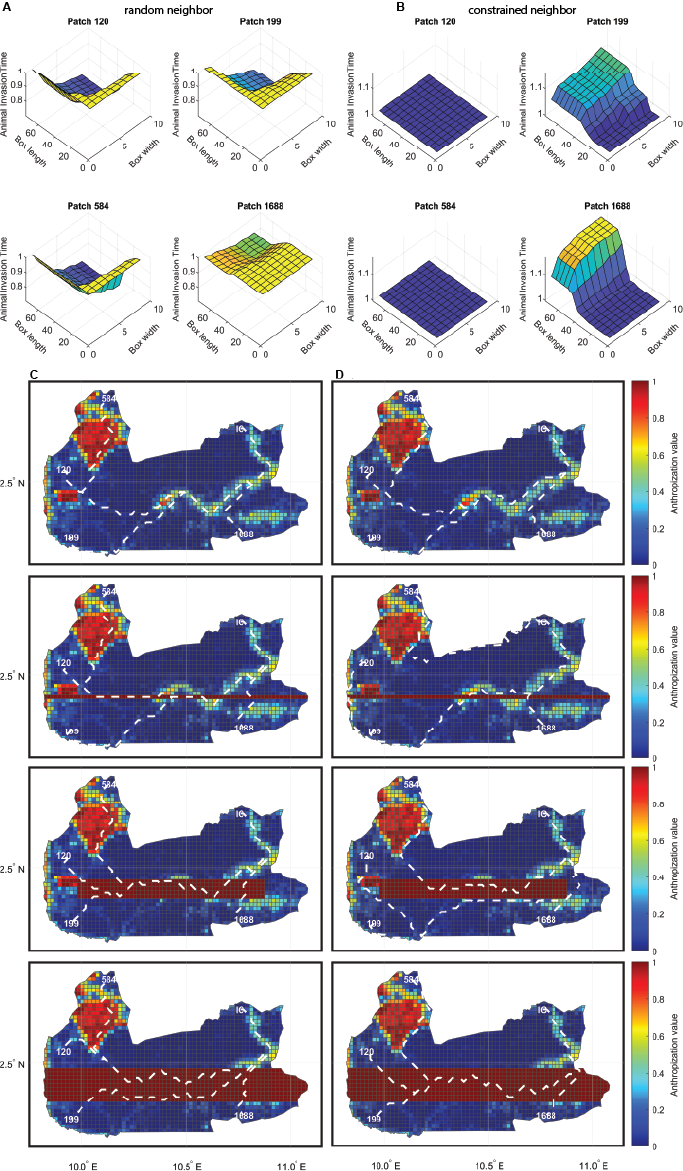}
    \caption{Invasion times and shortest paths between patches. \textbf{A,B}: A surface plot of invasion times to various patches is shown for domains that had an artificial rectangular region (size given in number of grid patches) of high anthropization introduced, centered at $2.39^\circ$ N, $10.427^\circ$ E. Invasion times are multiples of the invasion time in the original domain. \textbf{C,D}: Some example domains are shown alongside the shortest paths (in the sense of $t(i,j)$) from the location of the initial condition to the four patches at which invasion times were calculated. Panels A and C show random neighbor dispersal, panels B and D constrained neighbor dispersal; IC marks the patch in which the population was initialized.}
    \label{fig: invasion times}
\end{figure}

In addition to the steady state distributions, we investigated how different anthropization scenarios affected the speed of invasion. For this, we picked four patches within our domain that were a substantial distance away from the patch in which the population is initialized: a patch in the northwest (centered at $2.90^\circ$ N, $10.13^\circ$ E, 72 km from initial location, referred to as patch 584), a patch in the west (centered at $2.56^\circ$ N, $9.91^\circ$ E, 97 km from initial location, referred to as patch 120), a patch in the southwest (centered at $2.23^\circ$ N, $9.95^\circ$ E, 107 km from initial location, referred to as patch 199), and a patch in the southeast (centered at $2.25^\circ$ N, $10.78^\circ$ E, 56 km from initial location, referred to as patch 1688). In each simulation, we measured the time needed for the animal population to reach each of these four target patches. We defined invasion time as the first time the animal population exceeds half its eventual steady state density, i.e.
\begin{align*}
t_{\operatorname{inv},i} = \inf\left\{t: A_i(t)>\frac{A_i^*}{2}\right\}.
\end{align*}
Using different invasion thresholds did not yield qualitatively different invasion times (\cref{fig: appendix invasion theshold}). Comparison of invasion times across different simulations showed that the introduction of a highly anthropized area decreased invasion time in the case of random neighbor dispersal (\cref{fig: invasion times}A), while in the case of constrained neighbor dispersal, the presence of an anthropized region either increased invasion times or had a neutral effect (\cref{fig: invasion times}B). For random neighbor dispersal, invasion times even decreased for patches (120 and 584) for which the shortest path (in Euclidean distance) from the initial patch did not transect the artificially anthropized region. We thus concluded that population invasion does not follow the shortest path in Euclidean distance. Instead, we hypothesized that invasion takes place via a pulled traveling wave with Fisher speed. To test this hypothesis, we defined a new metric based on the time needed for a pulled traveling wave with Fisher speed. For this, we defined a weighted directed graph as follows. Nodes of the graph $N_i, i=1,\dots,n$ corresponded to the patches of our computational domain. Nodes had an edge connecting them if they were neighboring patches in the computational domain, i.e. $N_i$ and $N_j$ were connected in the direction from $N_i$ to $N_j$ if $i\in n_j$. Weights of edges represented the time needed for a population to invade $N_j$ from $N_i$ via a pulled traveling wave with Fisher speed. That is, we defined the weight for the edge connecting $N_i$ to $N_j$ as
\begin{align*}
w(N_i,N_j) = \frac{1}{2\sqrt{\left(r_{A,i}-\mu_{A,i}\right)D_{A}m_A(\alpha_{i})} },
\end{align*}
in the case of random neighbor dispersal and 
\begin{align*}
w(N_i,N_j) = \frac{1}{2\sqrt{\left(r_{A,i}-\mu_{A,i}\right)\beta^A_{[i\rightarrow j]}m_A(\alpha_{i}) }},
\end{align*}
in the case of constrained neighbor dispersal. Note that we used parameters governing animal dynamics due to the dependence of vector growth on animal density. Then, denoting by $\mathcal{P}(i,j):=\{P=p(\{1,\dots,K\}) \in \{N_1, \dots, N_n\}\}$, where $p:\{1,\dots,K\} \rightarrow \{N_1, \dots, N_n\}, p(1) = N_i, p(K)=N_j, K\in \mathbb{N}$ the set of all paths from $N_i$ to $N_j$, we defined the time required for a population to reach $N_j$ from $N_i$ via path $P\in \mathcal{P}(i,j)$ as
\begin{align*}
t_P(i,j) = \sum_{k=1}^{K-1} w(p(k),p(k+1)).
\end{align*}
Then, the time required for a population to invade $N_j$ from $N_i$ is 
\begin{align*}
t(i,j) = \min_{P\in \mathcal{P}(i,j)} t_P(i,j).
\end{align*}

Using this approximation for invasion times, we were able to explain the invasion time trends observed in \cref{fig: invasion times}A and B. In the case of random neighbor dispersal, we observed that the shortest path between patches in the sense of $t(i,j)$ was preferentially located within highly anthropized areas. This is because of the choice of dispersal term which increases emigration rates in highly anthropized areas, notwithstanding the lower carrying capacity (\cref{fig: invasion times}C). Thus, our artificially anthropized regions acted as ``highways'' for invasion, speeding up invasion times, even if the population did not have to cross the region (patches 120 and 584).

In the case of constrained neighbor dispersal, we observed that the shortest path between patches in the sense of $t(i,j)$ followed boundaries between areas of high and low anthropization (\cref{fig: invasion times}D). This is because movement speed along such boundaries is increased as the population preferentially moves into regions of low anthropization and thus, if a focal patch is surrounded by both patches of high and low anthropization, there is more movement into patches of low anthropization compared to another patch for which all neighbors feature low anthropization levels. As a consequence, shortest paths in this case avoided highly anthropized regions where possible (patches 120, 584), which explains the lack of impact on invasion times in these cases. By contrast, for patches that require migration through the artificial regions of high anthropization, the reduction of movement into these regions explains the increased invasion times.

\section{Conclusion-discussion}\label{section:conclusion}

This paper provides a  partially tractable framework connecting the landscape dynamics with those of wild animals and sylvatic hematophagous vectors. Designed as the opening stage of a comprehensive study on landscape-driven zoonotic emergence, it first maps out wild animal-vector interactions. 

Based on satellite images, the landscape `fragmentation' by anthropization is characterized and used in the model. Indeed, the model relies on a multi-patch framework in which each patch corresponds to a $2~\mathrm{km}\times 2~\mathrm{km}$ area \cite{Vale1984}, so as to mimic landscape heterogeneity. In each patch, the level of anthropization $\alpha$ is calculated using the land cover data.

Moreover, in each patch $i\in\{1,2,...,n\}$, it is assumed that
animals ($A_i$) are the only source of blood meals for hematophagous vectors $(V_i)$. This dependency is modeled with the saturating (Michaelis--Menten-type) function $\dfrac{A_i}{A_i+a_i}$ in the vector net growth term where $a_i$ is the half-saturation constant. In addition, the impact of anthropization level $\alpha_i$ is explicitly taken into account in the multi-patch model in the animal net growth rate term $g_i$ (see also \eqref{gi}), and in the between-patch migration terms of both animals and vectors. 
Indeed, we assume that the anthropization of the landscape leads to a loss of habitat for animals and induces a reduction of their carrying capacity. For simplicity, we follow \cite{dumont2025humanwildlifeinteractionstropicalforest} and assume that the animal carrying capacity in patch $i\in\{1,2,...,n\}$ is therefore $(1-\alpha_i)K_{A,i}$.
As a default assumption, we consider that any patch is a suitable habitat for both animals and vectors regardless of natural variation across patches. Therefore, the anthropization of patch $i\in\{1,2,...,n\}$, while reducing the available habitat (primarily for animals), may also induce the displacement of animals and vectors. Indeed, we assume that the
feedback of the anthropization level on the displacement rates 
has a sigmoidal shape with two branches: a low-displacement branch associated with low anthropization levels, and a high-displacement branch linked to high anthropization levels. Since on patch $i\in\{1,2,...,n\}$, animals $A_i$ are the only blood meal source for vectors $V_i$, we assume that the vector displacement rate out of patch $i$ is a decreasing function of $A_i$, to mimic the increased tendency of vectors to leave a patch when blood meals are scarce (low $A_i$). 

Altogether, these assumptions result in a nonlinear multi-patch system with nonlinear migration terms that extends the two-patch model studied in \cite{Happi2026}. 
Our model assumed that the emigration rate from a patch was a nonlinearly increasing function of patch anthropization. Our analytical results were derived for unconstrained global dispersal, in which individuals leaving a patch can move to any other patch of the domain. For the numerical simulations, we considered two more realistic formalisms: $(i)$ the case of random neighbor dispersal, and $(ii)$ the case of constrained neighbor dispersal. In case $(i)$, it is assumed that individuals prompted to leave patch $i\in\{1,2,...,n\}$ choose any neighboring patch at random. By contrast, in case $(ii)$ they preferably move to patches of low anthropization, quantified by the quality index $q_y$ (see \eqref{quality-index}). Here, $q_y$ is the proportion of emigrating animals that move into patch $y$; the higher $q_y$, the more likely animals are to choose $y$.

Furthermore, based on available literature, the multi-patch model was calibrated to the districts of Campo, Ny\'e'\'et\'e, and Ma'an (\cref{fig: fig1}A), located in the Ocean division (South Region, Cameroon) in the Guineo-Congolian region of the Biafran forest, a focus region of Human African Trypanosomiasis (HAT), commonly called sleeping sickness, and Animal African Trypanosomiasis (AAT), also known as nagana  \cite{Melachio_Tanekou2022-ox,Simo2014-oe}. HAT and AAT are neglected tropical diseases caused by the \textit{Trypanosoma} parasites for which tsetse flies (\textit{Glossina} spp.), hematophagous insects endemic to sub-Saharan Africa are the primary vectors. We first simulated the model with random neighbor dispersal of vectors and animals, i.e. system \eqref{system_red} with $q_y$ given by \eqref{q-random}. We observed that across the entire spatial domain, both animal and vector populations progressively expanded before ultimately settling into a stable equilibrium (\cref{fig: example sim unconstrained,fig: appendix ic tests}). By recording steady state densities ($A_i$ for animals and $V_i$ for vectors) in each patch $i\in\{1,2,...,n\}$, we found a linear relation between the vector steady state and the maximum growth rate of the vector population in patch $i$, namely $R_i:=r_{V,i}\frac{A_i}{A_i+a_i}$ (see also \eqref{fit-1} and \cref{fig: ss characterisation unconstrained}A). Moreover, we characterized the dependence of animal steady state density in patch $i$ on its anthropization level through the patch effective carrying capacity $K_{A,\operatorname{eff},i} = (1-\alpha_i)K_{A,i}$ and the per-capita emigration rate $\operatorname{em}_i := \frac{\alpha_i^\theta}{\alpha_i^\theta + c_A^\theta}$. A linear regression revealed a positive association between animal steady state density $A_i$ and animal effective carrying capacity. However, the low coefficient of determination ($R^2 \approx 0.32$) indicates a substantially nonlinear relationship (\cref{fig: ss characterisation unconstrained}B). In the same vein, with a linear regression model, we found that  $A_i$ and the emigration rate $\operatorname{em}_i$ exhibit an inverse relationship (\cref{fig: ss characterisation unconstrained}C).

We subsequently examined the constrained neighbor dispersal model \eqref{system_red}, applying the same numerical simulation framework to contrast its dynamics with random neighbor dispersal. Under constrained neighbor dispersal, animal and vector populations still progressively spread across the domain and ultimately reached a stable steady state (\cref{fig: example sim constrained,fig: appendix ic tests constrained}). However, compared with random neighbor dispersal, the rate of expansion was noticeably slower, and numerous patches maintained extremely sparse population levels. We re-evaluated the codependency between vector and host steady state densities. As shown in \cref{fig: ss characterisation constrained}A, vector density demonstrated a distinct piecewise linear relationship with the maximal vector growth rate $R_i$. Moreover, considering the central role played by the per-capita emigration rate (given by $\operatorname{em}_i$) under random neighbor dispersal, we subsequently analyzed how host equilibrium density varies as a function of $\operatorname{em}_i$. This interaction similarly exhibited a clear piecewise linear trend (see \cref{fig: ss characterisation constrained}B).

Moreover, we investigated how different anthropization scenarios
affected the speed of invasion. We observed that for random neighbor dispersal, invasion times decreased upon introductions of highly anthropized regions, even if the newly anthropized region was not along the shortest (in Euclidean sense) path between the population origin and target.
We thus concluded that population
invasion does not follow the shortest path in Euclidean distance. Approximating the invasion times based on pulled traveling wave with Fisher speed, we found that highly anthropized regions act as ``invasion highways''.
Under constrained neighbor dispersal, the shortest path between patches often traces the interface between high and low anthropization zones. This occurs because individuals travel faster along these edges as they actively favor moving into less anthropized environments. Consequently, optimal pathways skirt around heavily anthropized areas whenever feasible, accounting for why invasion times remain unaffected in those instances. Conversely, when migration directly through human-altered, high-anthropization zones is unavoidable, the lowered rate of movement into these areas drives the delay in invasion times.

A planned extension of the present contribution will add human demographics and a vector-borne pathogen to evaluate wild-to-human spillover dynamics mediated by habitat anthropization. Indeed, empirical and modeling studies suggest that environmental dynamics governing host and vector mobility reshape vector-borne disease epidemiology \cite{Crowder2013,Ferraguti2021,Fletcher2018,Lambin2010,Shaw2024,Tracey2014}, yet a notable shortage of both empirical and predictive modeling research exists to further clarify these processes.

\paragraph*{Data availability}
Computational code and data have been deposited in a BioStudies repository \cite{coderepo}.

\paragraph*{Acknowledgments:}
IT, IVY-D and PC acknowledge
the support of the AFRICAM project funded by the French Agency for Development (AFD) as part of PREACTS (PREZODE in Action in the Global South) program.

IVY-D and LE were supported by a Royal Society International Exchanges Award (IES\textbackslash R1\textbackslash 241236).

\paragraph*{Author contributions (using CRediT):}
O.W. Happi-Tchakount\'e: Methodology, Formal analysis, Writing - Original draft, Writing - Review \& Editing\\ 
I. Tcheferi: Methodology, Writing - Original draft, Writing - Review \& Editing\\
I.V. Yatat-Djeumen: Methodology, Formal analysis, Funding acquisition, Writing - Original draft, Writing - Review \& Editing, Supervision\\
L. Eigentler: Software, Validation, Formal analysis, Investigation, Data curation, Funding acquisition, Writing - Original draft, Writing - Review \& Editing, Visualization \\
P. Couteron: Methodology, Funding acquisition, Writing - Original draft, Writing - Review \& Editing

\paragraph*{Rights retention statement:}
For the purpose of open access, the authors have applied a Creative Commons Attribution (CC BY) licence to any Author Accepted Manuscript version arising from this submission.

\bibliographystyle{plain}
\bibliography{bibliography}

\appendix

\section{Results for an $n$-patch single-species equation}\label{AppendixD}
The inequalities between vectors are considered here in their usual coordinate-wise sense, i.e., for any $u=(u_i)_{i=1,...,p},\  v=(v_i)_{i=1,...,p}\in \mathbb{R}^p$, $p\geq1$,
\begin{itemize}
	\item $u\leq v \Leftrightarrow u_i\leq v_i$, $i=1,...,p$,
	\item $u<v \Leftrightarrow u\leq v$, $u\neq v$,
	\item $u\ll v \Leftrightarrow u_i<v_i,\ i=1,...,p$.
\end{itemize}

A square matrix $A=(a_{ij})$ is called a cooperative matrix if all its off-diagonal elements satisfy $a_{ij}\geq0$ for $i\neq j$. The spectrum of matrix $A$, written as $\sigma(A)$, is the set of eigenvalues of $A$.

\begin{definition}\label{def1}
	The stability modulus of a square matrix $A$, $s(A)$, is defined as
	\begin{align*}
s(A)=\max\{\operatorname{Re}\lambda: \lambda\in\sigma(A)\}.
\end{align*}
	
\end{definition}

\begin{definition}
	A square matrix $A$ is said to be reducible if there exists a permutation matrix $P$ such that $P^TAP$ is block triangular, where $P^T$ denotes the transpose matrix of $P$. Otherwise, $A$ is called irreducible.
\end{definition}

\begin{theorem}\label{Vidyasagar-Theorem} \cite{Dumont2012,vidyasagar_decomposition_1980}
	Consider the following $\mathcal{C}^1$ system
	\begin{equation}
	\def\arraystretch{2}
	\left\{ \begin{array}{l}
	\dfrac{dx}{dt} = f(x), \\
	\dfrac{dy}{dt} = g(x, y) 
	\end{array} \right.
	\label{equationVidyasagar}
	\end{equation}
	
	with $(x, y) \in \mathbb{R}^n \times\mathbb{R}^m$. Let $(x^*, y^*)$ be an equilibrium point.
	If $x^*$ is GAS in $\mathbb{R}^n$ for the system $\dfrac{dx}{dt} = f(x)$, and if $y^*$ is GAS in $\mathbb{R}^m$ for the system $\dfrac{dy}{dt} = g(x^*, y)$, then $(x^*, y^*)$ is (locally) asymptotically stable for system \eqref{equationVidyasagar}. Moreover, if all trajectories of \eqref{equationVidyasagar} are forward bounded, then $(x^*, y^*)$ is GAS for \eqref{equationVidyasagar}.
\end{theorem}

Consider the $n$-patch single-species dispersal model
\begin{equation}\label{n-patches-single-species}
\left\{
\begin{array}{l}
\dfrac{du_i}{dt}=u_iq_i(u_i)+\sum\limits_{j=1, j\neq i}^{n}D_{ij}u_j-\sum\limits_{{j=1, j\neq i}}^{n}D_{ji}u_i,\\
u_i(0)\geq0,
\end{array}	
\right.
\end{equation}
where $u_i(t)$ is a scalar representing the number of individuals in patch $i$ at time $t$, $q_i$ is the per-capita growth rate of the population in patch $i$, $D_{ij}$ is a non-negative migration coefficient from patch $j$ to patch $i$.
Consider also the connectivity matrix $D=(d_{ij})$ where
\begin{equation}\label{mat-d}
d_{ij}=\left\{\begin{array}{ll}
D_{ij}& \mbox{ if } i\neq j,\\
-\sum\limits_{k=1, k\neq i}^{n}D_{ki} & \mbox{ if } i= j.
\end{array}
\right.
\end{equation}
The Jacobian matrix of system \eqref{n-patches-single-species} at the origin is 
\begin{align*}
\mathcal{A}=D+\operatorname{diag}(q_i(0)).
\end{align*}
Finally, we introduce the hypotheses.
\begin{itemize}
	\item[(H1)] For any $i=1,2,3,...,n$, $q_i$ is decreasing on $\mathbb{R}_+$.
	\item[(H2)] For any $i=1,2,3,...,n$, $\lim\limits_{u_i\rightarrow+\infty}q_i(u_i)<0$.
	\item[(H3)] The matrix $D$ is irreducible.
\end{itemize}

The following results are valid.
\begin{theorem}\cite[Theorem 2.2]{Bliman2025-uf}\label{theorem:n-patches-single-species} Assume that (H1) holds.
	\begin{enumerate}
		\item If $s(\mathcal{A})\leq0$, then the origin is globally asymptotically stable (GAS) on $\mathbb{R}^n_+$ for system \eqref{n-patches-single-species}.
		\item If $s(\mathcal{A})>0$ and Assumptions (H2) and (H3) also hold, then the system \eqref{n-patches-single-species} admits a unique non-negative and non-zero equilibrium which is positive and GAS on $\mathbb{R}^n_+\setminus\{0_{\mathbb{R}^n}\}$.
	\end{enumerate}
	
\end{theorem}


\section{Appendix figures}

\begin{figure}
    \centering
    \includegraphics[height=0.85\textheight]{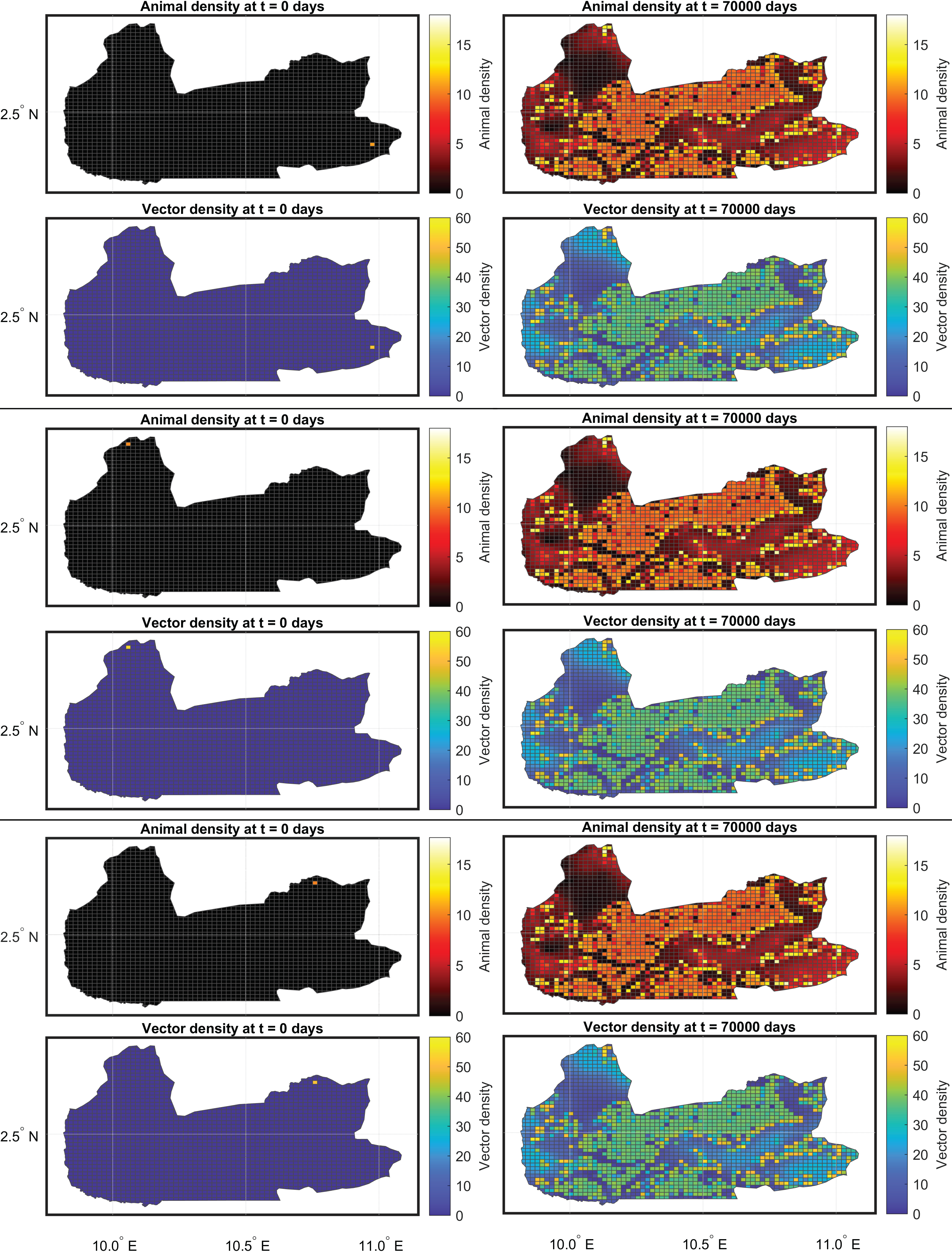}
    \caption{Simulations with different initial conditions show no difference in asymptotic model behavior. Three simulations of \eqref{system_red} (model with random neighbor dispersal) are shown. The difference between the three realizations is the initial condition (left column). Despite the different initial conditions, no different model outcomes (right column) are observed.}\label{fig: appendix ic tests}
\end{figure}

\begin{figure}
    \centering
    \includegraphics[height=0.85\textheight]{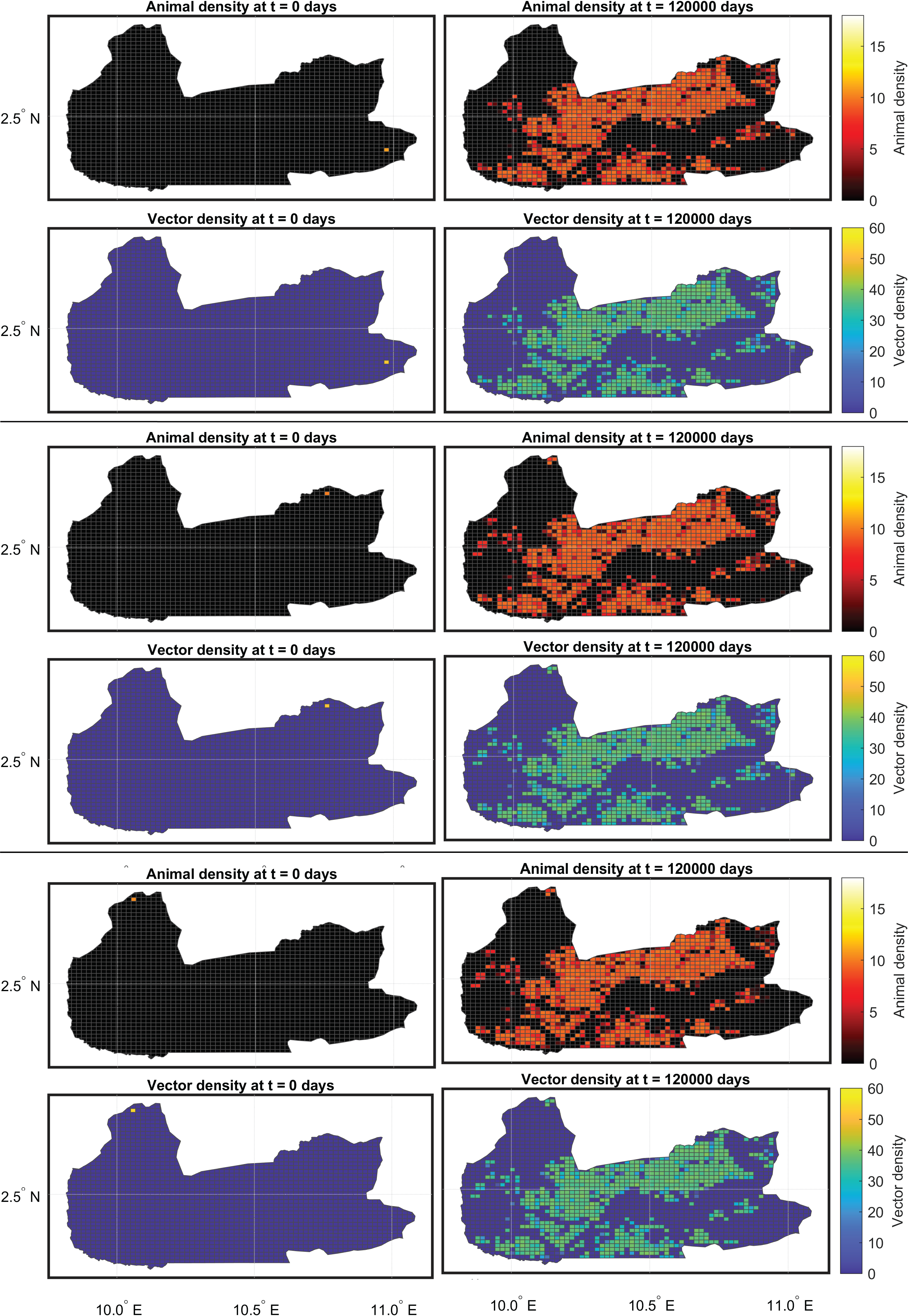}
    \caption{Simulations with different initial conditions show no difference in asymptotic model behavior. Three simulations of \eqref{system_red} (model with constrained neighbor dispersal) are shown. The difference between the three realizations is the initial condition (left column). Despite the different initial conditions, no different model outcomes (right column) are observed.}\label{fig: appendix ic tests constrained}
\end{figure}

\begin{figure}
    \centering
    \includegraphics[width=0.99\linewidth]{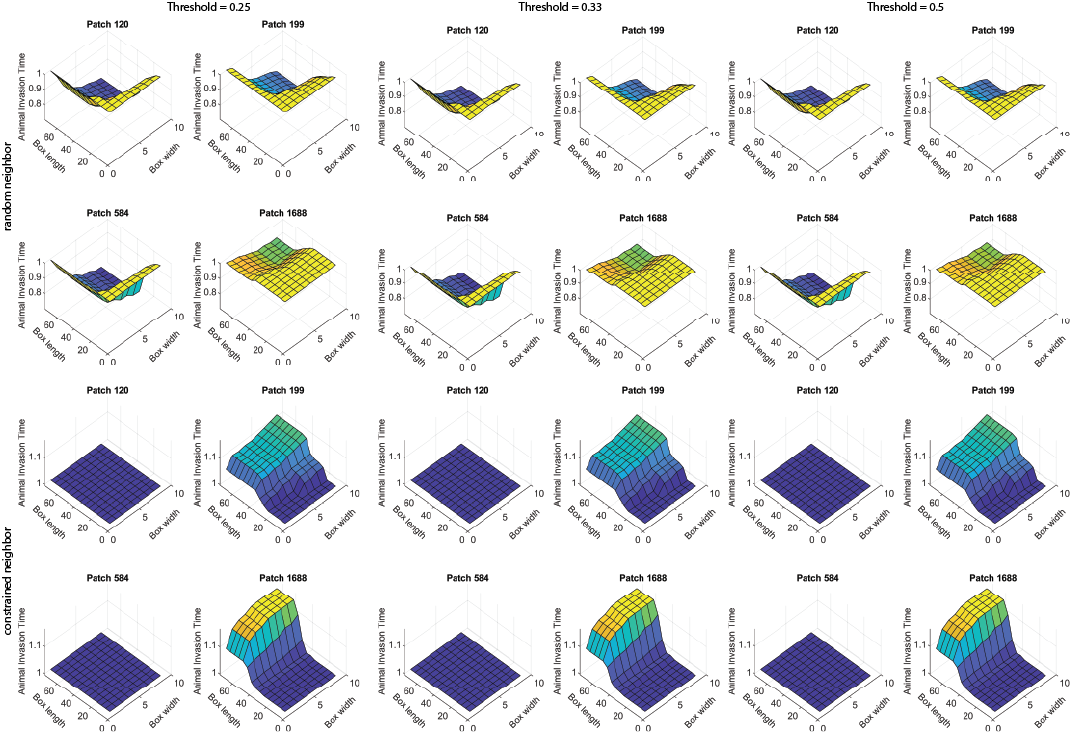}
    \caption{Invasion times do not depend qualitatively on the choice of invasion threshold. Invasion time surfaces described in \cref{fig: invasion times} are depicted for different choices of the invasion threshold.}
    \label{fig: appendix invasion theshold}
\end{figure}
\end{document}